\documentclass[article,a4paper,oneside,10pt]{memoir}

\usepackage{amsmath,amsthm}
\usepackage{amssymb}
\usepackage{amstext}
\usepackage{graphicx}
\usepackage{stmaryrd}

\usepackage{color}

\usepackage{pict2e}

\usepackage{xparse}

\makeatletter
\DeclareRobustCommand{\lintprod}{%
  \mathbin{\mathpalette\int@prod{(0,0)(0.8,0)(0.8,0.6)}}%
}
\DeclareRobustCommand{\rintprod}{%
  \mathbin{\mathpalette\int@prod{(0.1,0.6)(0.1,0)(0.9,0)}}}

\newcommand{\int@prod}[2]{%
  \begingroup
  \sbox\z@{$\m@th#1+$}%
  \setlength\unitlength{\wd\z@}%
  \linethickness{0.09\unitlength}%
  \begin{picture}(1,1)
  \roundcap
  \polyline#2
  \end{picture}%
  \endgroup
}
\makeatother

\usepackage[retainorgcmds]{IEEEtrantools}

\usepackage{tikz}

\newcommand{\Eb}{{\mathbf E}}
\newcommand{\Fb}{{\mathbf F}}

\newcommand{\Jb}{{\mathbf J}}

\newcommand{\Rb}{{\mathbf R}}

\newcommand{\Tb}{{\mathbf T}}
\newcommand{\ebf}{{\mathbf e}}

\newcommand{\fb}{{\mathbf f}}

\newcommand{\xb}{{\mathbf x}}

\newcommand{\ubf}{{\mathbf u}}
\newcommand{\vb}{{\mathbf v}}

\DeclareDocumentCommand \spinori { o } {%
  \IfNoValueTF {#1} {%
    \sigma%
  }{%
    \sigma_{#1}%
  }%
}

\DeclareDocumentCommand \indG { o o } {%
  \IfNoValueTF {#1} {%
    \iota%
  }{%
    \IfNoValueTF {#2} {%
      \iota_{#1}%
    }{%
      \iota_{#1}(#2)%
    }
  }%
}

\DeclareDocumentCommand \indGc { o o } {%
  \IfNoValueTF {#1} {%
    \bar\iota%
  }{%
    \IfNoValueTF {#2} {%
      \bar\iota_{#1}%
    }{%
      \bar\iota_{#1}(#2)%
    }
  }%
}

\DeclareDocumentCommand \coefG { o o } {%
  \IfNoValueTF {#1} {%
    \gamma%
  }{%
    \IfNoValueTF {#2} {%
      \gamma_{#1}%
    }{%
      \gamma_{#1}(#2)%
    }
  }%
}

\newcommand{\abf}{\mathbf{a}}
\newcommand{\bbf}{\mathbf{b}}

\newcommand{\wbf}{\mathbf{w}}

\renewcommand{\Fb}{\mf}

\newcommand{\semt}[1]{\mathbf{T}_\text{#1}}
\newcommand{\semti}[1]{T^\text{#1}}

\DeclareDocumentCommand \act { o } {%
  \IfNoValueTF {#1} {%
    \mathcal S%
  }{%
    \mathcal S_{\text{#1}}%
  }%
}

\DeclareDocumentCommand \ld { o } {%
  \IfNoValueTF {#1} {%
    \mathcal L%
  }{%
    \mathcal L_{\text{#1}}%
  }%
}

\newcommand{\drm}{\,\mathrm{d}}

\newcommand{\deltabf}{{\boldsymbol \partial}}

\newcommand{\xpert}[1][]{\varepsilon_{#1}}
\newcommand{\xbpert}{\pmb{\varepsilon}}

\newcommand{\mf}{{\mathbf F}}
\newcommand{\mfi}{F}

\newcommand{\sd}{{\mathbf J}}
\newcommand{\sdi}{J}

\newcommand{\vp}{{\mathbf A}}
\newcommand{\vpi}{{A}}

\newcommand{\gf}{{\mathbf G}}

\newcommand{\Gb}{\mathbf{G}}
\makeatletter
\def\trMs{\@ifnextchar[{\@with}{\@without}}
\def\@with[#1]{\Gb_{\xbpert}^{#1}}
\def\@without{\Gb_{\xbpert}^s}
\makeatother

\newcommand{\trcoeff}{G}

\DeclareMathOperator{\Tr}{Tr}

\DeclareMathOperator{\gr}{gr}

\newcommand{\len}[1]{\lvert#1\rvert}

\newtheoremstyle{question}
  {\topsep}   
  {\topsep}   
  {\upshape}  
  {0pt}       
  {\itshape}  
  {.}         
  {5pt plus 1pt minus 1pt} 
  {\thmname{#1} \thesection.\thmnumber{\itshape#2}\thmnote{(#3)}} 

\makeatletter
    \def\@endtheorem{\hfill$\P$\endtrivlist\@endpefalse }
\makeatother
\theoremstyle{question}

\makeatletter
\let\save@mathaccent\mathaccent
\newcommand*\if@single[3]{%
  \setbox0\hbox{${\mathaccent"0362{#1}}^H$}%
  \setbox2\hbox{${\mathaccent"0362{\kern0pt#1}}^H$}%
  \ifdim\ht0=\ht2 #3\else #2\fi
  }
\newcommand*\rel@kern[1]{\kern#1\dimexpr\macc@kerna}
\newcommand*\widebar[1]{\@ifnextchar^{{\wide@bar{#1}{0}}}{\wide@bar{#1}{1}}}
\newcommand*\wide@bar[2]{\if@single{#1}{\wide@bar@{#1}{#2}{1}}{\wide@bar@{#1}{#2}{2}}}
\newcommand*\wide@bar@[3]{%
  \begingroup
  \def\mathaccent##1##2{%
    \let\mathaccent\save@mathaccent
    \if#32 \let\macc@nucleus\first@char \fi
    \setbox\z@\hbox{$\macc@style{\macc@nucleus}_{}$}%
    \setbox\tw@\hbox{$\macc@style{\macc@nucleus}{}_{}$}%
    \dimen@\wd\tw@
    \advance\dimen@-\wd\z@
    \divide\dimen@ 3
    \@tempdima\wd\tw@
    \advance\@tempdima-\scriptspace
    \divide\@tempdima 10
    \advance\dimen@-\@tempdima
    \ifdim\dimen@>\z@ \dimen@0pt\fi
    \rel@kern{0.6}\kern-\dimen@
    \if#31
      \overline{\rel@kern{-0.6}\kern\dimen@\macc@nucleus\rel@kern{0.4}\kern\dimen@}%
      \advance\dimen@0.4\dimexpr\macc@kerna
      \let\final@kern#2%
      \ifdim\dimen@<\z@ \let\final@kern1\fi
      \if\final@kern1 \kern-\dimen@\fi
    \else
      \overline{\rel@kern{-0.6}\kern\dimen@#1}%
    \fi
  }%
  \macc@depth\@ne
  \let\math@bgroup\@empty \let\math@egroup\macc@set@skewchar
  \mathsurround\z@ \frozen@everymath{\mathgroup\macc@group\relax}%
  \macc@set@skewchar\relax
  \let\mathaccentV\macc@nested@a
  \if#31
    \macc@nested@a\relax111{#1}%
  \else
    \def\gobble@till@marker##1\endmarker{}%
    \futurelet\first@char\gobble@till@marker#1\endmarker
    \ifcat\noexpand\first@char A\else
      \def\first@char{}%
    \fi
    \macc@nested@a\relax111{\first@char}%
  \fi
  \endgroup
}
\makeatother

\newcounter{aside}
   {\refstepcounter{aside}
   \begin{adjustwidth}{\parindent}{-1.5\parindent}
	\small \textbf{Aside \thechapter.\theaside~#1}
	}
{\par\end{adjustwidth}}

\counterwithout{section}{chapter}
\settrimmedsize{297mm}{210mm}{*}
\settypeblocksize{664pt}{498.13pt}{*}
\setulmargins{3cm}{*}{*}
\setlrmargins{*}{*}{0.75}
\setmarginnotes{17pt}{51pt}{\onelineskip}
\setheadfoot{\onelineskip}{2\onelineskip}
\setheaderspaces{*}{2\onelineskip}{*}
\checkandfixthelayout

\providecommand{\keywords}[1]
{
  \small	
  \textbf{\textit{Keywords---}} #1
}

\usepackage{ulem}
\newcommand{\fieldA}{\abf}
\newcommand{\fieldB}{\bbf}
\newcommand{\fieldAi}{a}
\newcommand{\fieldBi}{b}
\newcommand{\abset}{\mathbf{T}_{\abf\cdot\bbf}}
\newcommand{\abseti}{T^{\abf\cdot\bbf}}
\newcommand{\Mat}{\mathbf{A}}
\newcommand{\Mati}{a}

\RequirePackage{latexsym}
\RequirePackage[colorlinks,citecolor=blue,urlcolor=blue,linkcolor=blue]{hyperref}

\begin{document}

\title{An exterior-algebraic derivation of the symmetric stress-energy-momentum tensor in flat space-time
\thanks{This work has been funded in part by the Spanish Ministry of Science, Innovation and Universities under grants TEC2016-78434-C3-1-R and BES-2017-081360. Published in: \textbf{\textit{Eur. Phys. J. Plus} 136, 212 (2021). DOI:} \href{https://doi.org/10.1140/epjp/s13360-021-01192-7}{10.1140/epjp/s13360-021-01192-7}.}
}

\author{\scshape Alfonso Martinez\thanks{alfonso.martinez@ieee.org},  Josep Font-Segura\thanks{josep.font@ieee.org}, Ivano Colombaro\thanks{ivano.colombaro@upf.edu} \thanks{Authors are with the Department of Information and Communication Technologies, Universitat Pompeu Fabra, Barcelona, Spain.}}

\maketitle

\begin{abstract}
This paper characterizes the symmetric rank-2 stress-energy-momentum tensor associated with fields whose Lagrangian densities are expressed as the dot product of two multivector fields, e. g.,  scalar or gauge fields, in flat space-time. The tensor is derived by a direct application of exterior-algebraic methods to deal with the invariance of the action to infinitesimal space-time translations; this direct derivation avoids the use of the canonical tensor. 
Formulas for the tensor components and for the tensor itself are derived for generic values of the multivector grade $s$ and of the number of time and space dimensions, $k$ and $n$, respectively. A simple, coordinate-free, closed-form expression for the interior derivative (divergence) of the symmetric stress-energy-momentum tensor is also obtained: this expression provides a natural generalization of the Lorentz force that appears in the context of the electromagnetic theory. Applications of the formulas derived in this paper to the cases of generalized electromagnetism, Proca action, Yang-Mills fields and conformal invariance are briefly discussed.
\end{abstract}



\section{Introduction: derivations of the symmetric stress-energy-momentum tensor}

The stress-energy-momentum (SEM) tensor, a symmetric, rank-2 tensor indexed by pairs of space-time indices, describes the flux of energy-momentum across regions of space-time; in a related manner, its divergence is used to express the conservation law for energy-momentum of isolated systems. In a different context, general relativity, the manifestly symmetric SEM tensor is also the source of the gravitational field. Both reasons amply support the importance of this tensor in mathematical physics. For electromagnetism, the SEM tensor combines the density of energy (component with time-time indices), the Poynting vector (components with space-time indices) and the Maxwell stress tensor (components with space-space indices) \cite[Sect.\ 33]{landau1982classicalTheoryFields}, \cite[Sect.\ 12.10]{jackson1999classicalElectrodynamics}. Moreover, the divergence of the SEM tensor is (minus) the Lorentz force \cite[Sect.\ 33]{landau1982classicalTheoryFields},\cite[Sect.\ 12.10]{jackson1999classicalElectrodynamics} that describes the effect of transfer of energy-momentum from the field to the charges.

In the context of field theories, the usual derivation of the SEM builds on the Lagrangian density and relates the invariance of the action with respect to infinitesimal space-time translations via Noether's theorem to the existence of a conserved current, which is in turn identified with the SEM tensor \cite{Forger_2004,Blaschke_2016}, \cite[Sect.\ 3.2]{maggiore2005modernIntroductionQFT}, \cite[Sect.\ 2.5]{diFrancesco1997conformalFieldTheory}. However, the canonical tensor obtained in this way is not necessarily symmetric and not necessarily gauge-independent for gauge theories. In order to symmetrize the canonical tensor, the Belinfante-Rosenfeld procedure is often used \cite{Forger_2004,Blaschke_2016}, \cite[Sect.\ 2.5]{diFrancesco1997conformalFieldTheory}. An alternative method that directly leads to a symmetric SEM tensor builds on the invariance of the action to variations in the Einstein-Hilbert space-time metric \cite{Forger_2004,Blaschke_2016}. In parallel, the study of generalized SEM tensors using the formalism of fiber bundles is a fertile area of research in mathematical physics \cite{gotay1992stressEnergyMomentumTensors,voicu2015energy-momentumTensors}.  

In this paper, we consider field theories whose Lagrangian densities can be expressed as linear combinations of the dot products of some pairs of multivector fields $\fieldA$ and $\fieldB$ of the same grade. We apply exterior-algebraic methods to deal with the invariance of the action of closed systems under infinitesimal space-time translations; Sect.~\ref{sec:exterior_algebra} provides a brief introduction to the necessary concepts of exterior algebra, including the formulation of the action for several field theories in exterior-algebraic terms. By means of a direct derivation that avoids the canonical tensor and yields a symmetric SEM tensor, we provide in Sect.~\ref{sec:SEM_tensor} several alternative, equivalent formulas for the SEM components and for the SEM tensor $\abset$ itself for generic values of the multivector grade $s$, 
\begin{equation}
	\abset = (-1)^{s}\bigl(\fieldA\odot\fieldB + \fieldA\owedge\fieldB\bigr),
\end{equation}
where the operations $\odot$ and $\owedge$ are defined in Sect.\ \ref{sec:tensor_fields}.
A simple, coordinate-free, closed-form expression for the interior derivative (divergence) of the SEM tensor, expressed in terms of the interior and exterior derivatives of multivector fields, respectively, denoted by $\deltabf\lintprod$ and $\deltabf\wedge$ and defined in Sect.\ \ref{sec:multivector_operations}, is also obtained: 
\begin{equation}
	\deltabf\lintprod\abset = \fieldA\lintprod(\deltabf\wedge\fieldB) + \fieldB\lintprod(\deltabf\wedge\fieldA) -\fieldA\rintprod(\deltabf\lintprod\fieldB)  - \fieldB\rintprod (\deltabf\lintprod\fieldA).
\end{equation}
This formula provides a natural generalization of the electromagnetic Lorentz force. Applications of the formulas derived in this paper to the cases of generalized electromagnetism (including scalar fields and electromagnetism), Proca action, Yang-Mills fields and conformal invariance are briefly discussed in Sect.~\ref{sec:applications}.

\section{Fundamentals of exterior algebra: notation and definitions} 
\label{sec:exterior_algebra}

\subsection{Multivectors}

We consider a flat space-time $\Rb^{k+n}$ with $k$ temporal dimensions and $n$ spatial dimensions. We represent the canonical basis of this $(k,n)$- or $(k+n)$-space-time by $\smash{\{\ebf_i\}_{i=0}^{k+n-1}}$; we adopt the convention that the first $k$ indices, i.~e.,~$i=0,\dotsc,k-1$, correspond to time components while the indices $i = k,\dotsc,k+n-1$ represent space components. Space and time have the same units and the speed of light is $c = 1$. Points and position in space-time are denoted by $\xb$, with components $x_i$ in the canonical basis $\smash{\{\ebf_i\}_{i=0}^{k+n-1}}$.

In exterior algebra, one considers vector field spaces whose basis elements $\ebf_I$ are indexed by lists $I = (i_1,\dotsc,i_m)$ drawn from $\mathcal{I}_m$, the set of all ordered lists with non-repeated $m$ indices, with $m \in \{0,1\dotsc,k+n$\}. We refer to elements of these vector field spaces as multivector fields of grade $m$. A multivector field $\fieldA(\xb)$ of grade $m$, possibly a function of the position $\xb$, with components $\fieldAi_I(\xb)$ in the canonical basis $\smash{\{\ebf_I\}_{I\in\mathcal{I}_m}}$ can be written as 
\begin{equation}\label{eq:multivector}
\fieldA(\xb) = \sum_{I\in\mathcal{I}_m} \fieldAi_I(\xb)\ebf_I.
\end{equation}
The dimension of the vector space of all grade-$m$ multivectors is $\smash{\binom{k+n}{m}}$, the number of lists in $\mathcal{I}_m$. We denote by $\len{I}$ the length of a list $I$ and 
by $\gr(\fieldA)$ the operation that returns the grade of a vector $\fieldA$.

\subsection{Operations on multivectors}
\label{sec:multivector_operations}

We next define several operations acting on multivectors; our presentation loosely follows \cite[Sect.~2 and~3]{colombaro2019introductionSpaceTimeExteriorCalculus} and \cite[Sect.~2]{colombaro2020generalizedMaxwellEquations}. Introductions to exterior algebra from the perspective and language of differential forms can be found in \cite{lovelock1989tensorsDifferentialForms}, \cite{flanders1989differentialForms}.
With no real loss of generality, we define the operations only for the canonical basis vectors, the operation acting on general multivectors being a mere extension by linearity of the former. 
First, the dot product $\cdot$ of two arbitrary grade-$m$ basis vectors $\ebf_I$ and $\ebf_J$ is defined as
\begin{equation}\label{eq:dot_multi}
	\ebf_I\cdot\ebf_J = \Delta_{IJ} = \Delta_{i_1 j_1}\Delta_{i_2 j_2}\dotsm\Delta_{i_m j_m},
\end{equation}	
where $I$ and $J$ are the ordered lists $I = (i_1,i_2,\dotsc,i_m)$ and $J = (j_1,j_2,\dotsc,j_m)$ and $\Delta_{ij} = 0$ if $i\neq j$, and we let time unit vectors $\ebf_{i}$ have negative metric $\Delta_{ii} = -1$ and space unit vectors $\ebf_{i}$ have positive metric $\Delta_{ii} = +1$. 

Let two basis vectors $\ebf_I$ and $\ebf_J$ have grades $m = \len{I}$ and $m' = \len{J}$. Let $(I,J) = \{i_1,\dotsc,i_m,j_1,\dotsc,j_{m'}\}$ be the concatenation of $I$ and $J$, let $\sigma(I,J)$ denote the signature of the permutation sorting the elements of this concatenated list of $\len{I}+\len{J}$ indices, and let $I+J$, or $\varepsilon(I,J)$ if the notation $I+J$ is ambiguous in a given context, denote the resulting sorted list. For a generic list $I$, let $\sigma(I)$ denote the signature of the permutation sorting the elements of the list $I$.
Then, the exterior product of $\ebf_I$ and $\ebf_J$ is defined as
\begin{equation} \label{eq:ext-prod-def}
	\ebf_I\wedge\ebf_J = \sigma(I,J)\ebf_{I+J}.
\end{equation}
The exterior product is thus either zero or a vector of grade $\len{I}+\len{J}$. The exterior product provides a construction of the basis vector $\ebf_I$, with $I$ an ordered list $I = (i_1,\dotsc,i_m)$,  from the canonical basis vectors $\ebf_i$, namely
\begin{equation}
	\ebf_I = \ebf_{i_1}\wedge\ebf_{i_2}\wedge\dotsm\wedge\ebf_{i_m}.
\end{equation}

We next define two generalizations of the dot product, the left and right interior products. 
Let $\ebf_I$ and $\ebf_J$ be two basis vectors of respective grades $\len{I}$ and $\len{J}$. The left interior product, denoted by $\lintprod$, is defined as
\begin{equation}
	\ebf_I \lintprod \ebf_J = \Delta_{II}\sigma(J\setminus I,I)\ebf_{J\setminus I},
	\label{eq:left-int-prod}
\end{equation}
if $I$ is a subset of $J$ and zero otherwise. The vector $\ebf_{J\setminus I}$ has grade $\smash{\len{J}-\len{I}}$ and is indexed by the elements of $J$ not in common with $I$.
The right interior product, denoted by $\rintprod$, of two basis vectors $\ebf_I$ and $\ebf_J$ is defined as 
\begin{equation}
	\ebf_I \rintprod \ebf_J = \Delta_{JJ}\sigma(J,I\setminus J)\ebf_{I\setminus J}, \label{eq:right-int-prod}
\end{equation}
if $J$ is a subset of $I$ and zero otherwise.
Both interior products are grade-lowering operations, as the left (resp.~right) interior product is either zero or a multivector of grade $\smash{\len{J}-\len{I}}$ (resp.~$\len{I}-\len{J}$).

We define the differential vector operator $\deltabf$ as $(-\partial_0,-\partial_2,\dotsc,-\partial_{k-1}, \partial_{k},\dotsc,\partial_{k+n-1})$, that is
\begin{equation}
	\deltabf = \sum_{i=0}^{k+n-1}\Delta_{ii}\ebf_i\partial_i.
\end{equation}	
As done in \cite[Sect.~3]{colombaro2019introductionSpaceTimeExteriorCalculus}, we define the exterior derivative of $\vb$, $\deltabf\wedge \vb$, of a given vector field $\vb$ of grade $m$ as
\begin{equation}\label{eq:ext-deriv-def}
\deltabf\wedge \vb = \sum_{i,I\in\mathcal{I}_m:\, i\notin I} \Delta_{ii}\sigma(i,I)\partial_i v_I \ebf_{i+I}.
\end{equation}
In addition, we define the interior derivative of $\vb$ as $\deltabf\lintprod \vb$, namely
\begin{equation}\label{eq:int-deriv-def}
\deltabf\lintprod \vb = 
\sum_{i,I\in\mathcal{I}_m:\, i\in I} \sigma(I\setminus i,i) \partial_i v_I \ebf_{I\setminus i}.
\end{equation}
The exterior and interior derivatives satisfy the property $\deltabf\wedge(\deltabf\wedge \vb) = 0 = \deltabf\lintprod(\deltabf\lintprod\vb)$.

\subsection{Exterior-algebraic formulation of the Lagrangian densities and the action}
\label{sec:Lagrangian_exterior_algebra}

We review how several Lagrangian densities may be written as the linear combination of dot products of pairs of multivector fields. First, the free Lagrangian density for generalized electromagnetism \cite{colombaro2020generalizedMaxwellEquations} is given by
\begin{align}\label{eq:lagrangian_gem_free}
	\ld_\text{free-gem} &= \frac{(-1)^{r-1}}{2}\mf\cdot\mf,
\end{align}
where $\mf$ is a grade-$r$ multivector field related to the generalized potential $\vp$, a multivector field of grade $r-1$, as $\mf = \deltabf\wedge\vp$. This is a gauge theory, since replacing the potential $\vp$ by a new field $\vp' = \vp + \bar\vp + \deltabf\wedge\gf$, where $\bar\vp$ is a constant $(r-1)$-vector and $\gf$ is an $(r-2)$-vector gauge field, leaves $\mf$  and therefore the Lagrangian unchanged. 

The model in \eqref{eq:lagrangian_gem_free} subsumes a number of relevant cases. Setting $r = 1$, $k = 1$, $n = 3$, and $\vp = \phi$, gives the scalar field Lagrangian \cite[Sect.\ 3.3]{maggiore2005modernIntroductionQFT}, \cite{Forger_2004,Blaschke_2016},
\begin{align}\label{eq:lagrangian_scalar_free}
	\ld_\text{free-scalar} &= \frac{1}{2}(\deltabf\wedge \phi)\cdot(\deltabf\wedge \phi) \\
	&= \frac{1}{2}\sum_{i}\Delta_{ii}(\partial_i\phi)^2.
\end{align}

For electrostatics, we set $r = 1$, $k = 0$, $n =3$, and the Lagrangian density \cite[Ch.\ 19]{feynman1977lecturesPhysicsvol2} is given by    
\begin{equation}\label{eq:lagrangian_electrostatics_free}
\ld_\text{free-es} = \frac{1}{2}\Eb\cdot\Eb,
\end{equation}
where $\Eb$ is the electric field and $\phi$ is the opposite in sign of the usual electric potential, that is $\Eb = \deltabf \wedge \phi = \nabla \phi$. For electromagnetism, we set $r = 2$, $k = 1$, $n =3$; in this case, when the Lagrangian is expressed in terms of the Faraday tensor, an antisymmetric tensor of rank 2, the factor before $\mf\cdot\mf$ becomes $-\frac{1}{4}$ to account for the repeated sum over pairs of indices \cite[Sect.\ 27]{landau1982classicalTheoryFields}, \cite[Sect.\ 12.10]{jackson1999classicalElectrodynamics},
\cite[Sect.\ 3.5]{maggiore2005modernIntroductionQFT}, \cite[Sect.\ 1.5]{zee2003quantumFieldTheoryNutshell}. We have instead,
\begin{equation}\label{eq:lagrangian_em_free}
	\ld_\text{free-em} = -\frac{1}{2}\mf\cdot\mf.
\end{equation}

In Yang-Mills theories in Minkowksi space-time, the multivector field takes values in a Lie algebra, and the free Lagrangian density \cite[Sect.\ 10.2]{maggiore2005modernIntroductionQFT}, \cite{Forger_2004,Blaschke_2016}, is given by
\begin{align}
	\ld_\text{free-ym} &= -\frac{1}{2g^2}\Tr(\mf\cdot\mf) \label{eq:lagrangian_ym_free_1} \\
	&= -\frac{1}{2g^2}\sum_a \mf^a\cdot\mf^a, \label{eq:lagrangian_ym_free_2}
\end{align}
where $g$ is a coupling parameter, \eqref{eq:lagrangian_ym_free_1} is the fundamental representation and \eqref{eq:lagrangian_ym_free_2} the form as an linear combination of real-valued multivector fields in a given representation of the Lie algebra.

In addition to the free Lagrangian density, the action may include interaction terms. 
For instance, we may have a source density, a vector $\sd(\xb)$ of grade $(r-1)$ \cite{colombaro2020generalizedMaxwellEquations}, such that the  generalized-electromagnetism  Lagrangian density \cite[Sect.\ 12.7]{jackson1999classicalElectrodynamics}, is now
\begin{align}\label{eq:lagrangian_gem}
	\ld_\text{gem} &= \frac{(-1)^{r-1}}{2}\mf\cdot\mf + \sd\cdot\vp.
\end{align}
For this density, the requirement that the action is gauge invariant leads to the continuity equation for the current density, $\deltabf\lintprod\sd = 0$. As shown in \cite[Sect.\ 4]{colombaro2020generalizedMaxwellEquations}, the generalized Lorentz force density remains a grade-1 vector, whose value coincides with the opposite in sign of the divergence of the stress-energy-momentum tensor related to \eqref{eq:lagrangian_gem}.

Alternatively, the free Lagrangian density in \eqref{eq:lagrangian_em_free} may describe a generalized Proca equation \cite[Sect.\ 12.8]{jackson1999classicalElectrodynamics}, \cite[p.\ 107]{maggiore2005modernIntroductionQFT} with mass $m$ and Lagrangian density
\begin{equation}\label{eq:lagrangian_proca}
	\ld_\text{free-proca} = -\frac{1}{2}\mf\cdot\mf + \frac{1}{2}m^2\vp\cdot\vp.
\end{equation}

In each of the cases we have described, namely \eqref{eq:lagrangian_scalar_free}, \eqref{eq:lagrangian_electrostatics_free}, \eqref{eq:lagrangian_ym_free_1}, \eqref{eq:lagrangian_gem} and \eqref{eq:lagrangian_proca}, one can construct an action $\act_\text{sys}$ from the Lagrangian densities by integrating the density over a region $\mathcal{R}$ that comprises the system under consideration. 
We shall assume that the region $\mathcal{R}$ is large enough to make the physical system closed, so that an infinitesimal translation of space-time coordinates is a symmetry of the system and the action is invariant under this translation. We shall also assume that the fields decay fast enough over $\mathcal{R}$ so that the flux of the fields over the boundary of $\mathcal{R}$ is arbitrarily small.
In general, we may write the action $\act_\text{sys}$ as
\begin{equation}\label{eq:action-00a}
	\act_\text{sys} = \int_{\mathcal{R}}\! \drm^{k+n}\xb\, \Biggl(\sum_{\fieldA,\fieldB}\gamma_{\fieldA,\fieldB}\, (\fieldA\cdot\fieldB)\Biggr),
\end{equation} 
namely an integral over $\mathcal{R}$ of a linear combination of dot products of properly selected pairs of multivector fields $\fieldA$ and $\fieldB$ of the same degree, with respective coefficients $\gamma_{\fieldA,\fieldB}$. 
We exploit this linearity to express the stress-energy momentum tensor associated with the action \eqref{eq:action-00a} as a linear combination of tensors $\abset$, as we shall see in Sect.\ \ref{sec:SEM_tensor}.

\subsection{Symmetric tensor field spaces}
\label{sec:tensor_fields}

In addition to exterior-algebraic vector spaces, we also need to consider symmetric tensor field (vector) spaces and matrix field (vector) spaces. The symmetric tensor field space has basis elements $\ubf_{J}$ indexed by lists $J = (j_1,\dotsc,j_r)$ drawn from $\mathcal{J}_r$, the set of all ordered lists of (possibly repeated) $r$ indices. We refer to elements of these vector spaces as symmetric tensors of rank $r$.
The dimension of the vector space of rank-$r$ symmetric tensors is $\smash{\binom{k+n+r-1}{r}}$, the number of lists in $\mathcal{J}_r$. The dot product $\cdot$ of two arbitrary rank-$r$ basis vectors $\ubf_I$ and $\ubf_J$ is defined as
\begin{equation}\label{eq:dot_symmtens}
	\ubf_I\cdot\ubf_J = \Delta_{IJ} = \Delta_{i_1 j_1}\Delta_{i_2 j_2}\dotsm\Delta_{i_r j_r},
\end{equation}	
where $I$ and $J$ are the ordered lists $I = (i_1,i_2,\dotsc,i_r)$ and $J = (j_1,j_2,\dotsc,j_r)$. Let two basis vectors $\ubf_I$ and $\ubf_J$ have respective grades $m = \len{I}$ and $m' = \len{J}$. Then, the symmetric tensor product of $\ubf_I$ and $\ubf_J$ is defined as
\begin{equation} \label{eq:symm-prod-def}
	\ubf_I\vee\ubf_J = \ubf_{I+J}.
\end{equation}
We may thus construct the basis vector $\ubf_I$, with $I \in \mathcal{J}_r$, from the canonical basis vectors elements $\ebf_i$ as
\begin{equation}
	\ubf_I = \ebf_{i_1}\vee\ebf_{i_2}\vee\dotsm\vee\ebf_{i_r}.
\end{equation}
We may also define the interior product between a multivector and a symmetric tensor. Let $\ebf_I$ and $\ubf_J$ be two basis vectors of respective grade $\len{I}$ and rank $\len{J}$. The interior product, indistinctly denoted by $\lintprod$ or $\rintprod$, is defined as
\begin{equation}
	\ebf_I \lintprod \ubf_J = \ubf_J \rintprod \ebf_I = \Delta_{II}\ubf_{J\setminus I},
	\label{eq:left-int-prod-multiv-tens}
\end{equation}
if $I$ is a subset of $J$ and zero otherwise. In this case, left and right interior products coincide.

We shall need to consider only rank-2 symmetric tensors, as the stress-energy-momentum tensor is a symmetric tensor field of rank 2, irrespective of the grade of the multivector fields and the number of space dimensions. In this case, it will prove convenient to have two operations, $\owedge$ and $\odot$, that generate a rank-2 symmetric tensor from a pair of multivectors of the same grade, $\fieldA$ and $\fieldB$. These operations are, respectively, defined as
\begin{gather}
	\fieldA\owedge\fieldB = \sum_{i\leq j}(\Delta_{ii}\ebf_i\wedge\fieldA)\cdot(\fieldB\wedge\ebf_j\Delta_{jj})\ubf_{ij}, \label{eq:def-owedge} \\
	\fieldA\odot\fieldB = \sum_{i\leq j}(\Delta_{ii}\ebf_i\lintprod\fieldA)\cdot(\fieldB\rintprod\ebf_j\Delta_{jj})\ubf_{ij}. \label{eq:def-odot}
\end{gather}
While the tensor field basis $\ubf_{ij}$ is symmetric in the pair $(i,j)$, that is $\ubf_{ij} = \ubf_{ji}$, the tensors defined in \eqref{eq:def-owedge} and \eqref{eq:def-odot} are not separately symmetric, as the coefficients multiplying $\ubf_{ij}$, for $i\neq j$, are not symmetric over the pair $(i,j)$. However, for $i \neq j$, from \cite[Eq.\ (22)]{colombaro2020generalizedMaxwellEquations}, we have that 
\begin{equation}
	(\Delta_{ii}\ebf_i\wedge\fieldA)\cdot(\fieldB\wedge\ebf_j\Delta_{jj}) = (\Delta_{jj}\ebf_j\lintprod\fieldA)\cdot(\fieldB\rintprod\ebf_i\Delta_{ii}),
\end{equation}
and therefore the $(i,j)$ coefficient of $\fieldA\owedge\fieldB$, with $i< j$ (resp.\ $j >i$), coincides with the value of the coefficient $(j,i)$ of $\fieldA\odot\fieldB$, with $j > i$ (resp.\ $j < i$).
Therefore, the sum tensor $\fieldA\owedge\fieldB + \fieldA\odot\fieldB$ is indeed symmetric over the pair $(i,j)$.

\subsection{Matrix vector spaces}

We do not need consider general tensor fields, but rather the matrix field (vector) space whose basis elements can be represented as $\wbf_{I_1,I_2}$, where both $I_1$ and $I_2$ are ordered lists of non-repeated $\ell_1$ and $\ell_2$ elements, respectively. We may identify these basis elements with the tensor product of two multivectors of grade $\ell_1$ and $\ell_2$, namely
\begin{equation}
	\wbf_{I_1,I_2} = \ebf_{I_1}\otimes\ebf_{I_2}.
\end{equation}
The dimension of the vector space spanned by these basis elements is $\smash{\binom{k+n}{\ell_1}\binom{k+n}{\ell_2}}$; the elements of this vector space can be identified with matrices $\Mat$ whose rows and columns are indexed by lists of $I_i \in \mathcal{I}_{\ell_i}$, 
\begin{equation}
	\Mat = \sum_{I_1\in\mathcal{I}_{\ell_1},I_2\in\mathcal{I}_{\ell_2}}\Mati_{I_1I_2}\wbf_{I_1,I_2}. 
\end{equation}
The transpose of a matrix element $\wbf_{I_1,I_2}$, denoted as $\wbf_{I_1,I_2}^T$, is defined as $\wbf_{I_2,I_1}$. For $\ell_1 = \ell_2 = \ell$, we also define the grade-$\ell$ square identity matrix, denoted by $\mathbf{1}_\ell$, as
\begin{equation}\label{eq:identity-matrix-m}
	\mathbf{1}_\ell = \sum_{I\in\mathcal{I}_\ell}\Delta_{II}\wbf_{I,I}. 
\end{equation}

The dot product $\cdot$ of two arbitrary matrix elements $\wbf_{I_1,I_2}$ and $\wbf_{J_1,J_2}$ is defined as
\begin{equation}\label{eq:dot_tensor_general}
	\wbf_{I_1,I_2}\cdot\wbf_{J_1,J_2} = \Delta_{I_1J_1}\Delta_{I_2J_2}.
\end{equation}	
We also define the dot product between symmetric tensor and matrix basis elements, $\ubf_I$ and $\wbf_{J_1,J_2}$, as
\begin{equation}\label{eq:dot_tensor_symm}
	\ubf_I \cdot \wbf_{J_1,J_2} =  \wbf_{J_1,J_2} \cdot \ubf_I = \begin{cases} \Delta_{I,I}, \quad &\text{ if } I = \varepsilon(J_1,J_2), \\
	0, \quad &\text{ otherwise.}	
	 \end{cases}
\end{equation}
We define the matrix product $\times$ between two matrix basis elements $\wbf_{I,J}$ and $\wbf_{K,L}$ as
\begin{equation}\label{eq:matrix_vector}
	\wbf_{I,J} \times \wbf_{K,L} =  \wbf_{I,L} \Delta_{JK}.
\end{equation}
For square matrices $\Mat$ indexed by grade-$m$ multivectors, it is natural to define the matrix inverse, denoted as $\Mat^{-1}$, as the matrix such that $\Mat^{-1}\times\Mat = \mathbf{1}_m = \Mat\times\Mat^{-1}$.
And lastly, we define the matrix product $\times$ between a matrix $\wbf_{I,J}$ and a multivector $\ebf_K$ (or between a multivector $\ebf_K$ and the matrix $\wbf_{J,I}$, i.e., the transpose of $\wbf_{I,J}$) as
\begin{equation}\label{eq:matrix_vector_transp}
	\wbf_{I,J} \times \ebf_K =  \ebf_K \times \wbf_{J,I} =  \ebf_{I} \Delta_{JK}.
\end{equation}

\section{Variational derivation of the symmetric energy-momentum tensor as conserved current with respect to infinitesimal space-time translations}
\label{sec:SEM_tensor}

\subsection{Introduction}

In all the systems described in Sect.\ \ref{sec:Lagrangian_exterior_algebra}, namely \eqref{eq:lagrangian_scalar_free}, \eqref{eq:lagrangian_electrostatics_free}, \eqref{eq:lagrangian_ym_free_1}, \eqref{eq:lagrangian_gem} and \eqref{eq:lagrangian_proca}, we could construct an action $\act_\text{sys}$ by integrating the Lagrangian density over a region $\mathcal{R}$ that comprises the physical system under consideration.  As we stated earlier, we assume that the region $\mathcal{R}$ is large enough to make the physical system closed, so that an infinitesimal translation of space-time coordinates is a symmetry of the system and the action is invariant under this change. We also assume that the fields decay fast enough over $\mathcal{R}$ so that the flux of the fields over the boundary of $\mathcal{R}$ is arbitrarily small. The action for the physical system under consideration, $\act_\text{sys}$ was written in \eqref{eq:action-00a} as
\begin{equation}\label{eq:action-0a}
	\act_\text{sys} = \int_{\mathcal{R}}\! \drm^{k+n}\xb\, \Biggl(\sum_{\fieldA,\fieldB}\gamma_{\fieldA,\fieldB}\, (\fieldA\cdot\fieldB)\Biggr),
\end{equation} 
namely an integral over $\mathcal{R}$ of a linear combination of dot products of properly selected pairs of multivector fields $\fieldA$ and $\fieldB$ of the same degree, with respective coefficients $\gamma_{\fieldA,\fieldB}$. We exploit this linearity to express the stress-energy momentum tensor associated with the action \eqref{eq:action-0a} as a linear combination, with coefficients $\gamma_{\fieldA,\fieldB}$, of tensors $\abset$ associated with Lagrangian densities $\fieldA\cdot\fieldB$, namely
\begin{equation}\label{eq:sem_total}
	\semt{sys} = \sum_{\fieldA,\fieldB}\gamma_{\fieldA,\fieldB}\abset,
\end{equation}
where the tensor $\abset$ is given by $\abset = (-1)^{s}\bigl(\fieldA\odot\fieldB + \fieldA\owedge\fieldB\bigr)$ in \eqref{eq:abset-oo}, with components given in \eqref{eq:abset_ii} and \eqref{eq:abset_ij},
\begin{gather}
	\abseti_{ii} = \Delta_{ii}\Biggl(\sum_{K\in{\mathcal{I}_{s}}:i\notin K}\Delta_{KK}\fieldAi_{K}\fieldBi_{K}-\sum_{K\in{\mathcal{I}_{s}}:i\in K}\Delta_{KK}\fieldAi_{K}\fieldBi_{K}\Biggr), \label{eq:abset_ii_intro} \\
	\abseti_{ij} = -\sum_{K\in{\mathcal{I}_{s-1}}:i,j\notin K}\Delta_{KK}\sigma\bigl(\varepsilon(i,K)_{i\leftrightarrow j}\bigr)\bigl(\fieldAi_{\varepsilon(i,K)}\fieldBi_{\varepsilon(j,K)}+\fieldBi_{\varepsilon(i,K)}\fieldAi_{\varepsilon(j,K)}\bigr), \label{eq:abset_ij_intro}
\end{gather}
where $I_{i\leftrightarrow j}$ is a list of indices where the index $i$ in $I$ is replaced by $j$, for a generic $I$. 
As we shall see, consideration of the possible internal structure of the fields $\fieldA$ or $\fieldB$, e. g., $\fieldA = \deltabf\wedge\vp$ in generalized electromagnetism or an analogous formula for the Yang-Mills fields, may be circumvented to directly obtain the tensor components \eqref{eq:abset_ii_intro}--\eqref{eq:abset_ij_intro}  in terms of the components of the fields $\fieldA$ and $\fieldB$ appearing in the Lagrangian density.

Moreover, we prove in \eqref{eq:int-der-abset-cf} that the interior derivative (divergence) of $\semt{sys}$ is given by
\begin{equation}
	\deltabf\lintprod\semt{sys} = \sum_{\fieldA,\fieldB}\gamma_{\fieldA,\fieldB}\,\bigl(\fieldA\lintprod(\deltabf\wedge\fieldB) + \fieldB\lintprod(\deltabf\wedge\fieldA) -\fieldA\rintprod(\deltabf\lintprod\fieldB)  - \fieldB\rintprod (\deltabf\lintprod\fieldA)\bigr), \label{eq:int-der-abset-cf-intro}
\end{equation}
in terms of interior and exterior derivatives of $\fieldA$ and $\fieldB$. Assuming that infinitesimal space-time translations are a symmetry of the system under consideration and that the fields decay sufficiently fast at the boundary of the region $\mathcal{R}$, this interior derivative is zero, 
\begin{equation}
	\deltabf\lintprod\semt{sys} = 0,
\end{equation}
which yields a conservation law for the energy-momentum of the system.

\subsection{Action for the Lagrangian density $\fieldA\cdot\fieldB$}

The linearity of the action in \eqref{eq:action-1a} allows us to find the tensor and its interior derivative as a linear combination of quantities derived from the Lagrangian density $\fieldA\cdot\fieldB$ and its action $\act_{\fieldA\cdot\fieldB}$, given by
\begin{equation}\label{eq:action-1a}
	\act_{\fieldA\cdot\fieldB} = \int_{\mathcal{R}}\! \drm^{k+n}\xb\, (\fieldA\cdot\fieldB).
\end{equation}

Let us shift the origin of coordinates by an infinitesimal perturbation $\xbpert$ and denote by $\{\ebf\}$ and $\{\ebf'\}$, respectively, the original and shifted basis elements, both expressed in the original basis. In general, and with some abuse of notation, we denote the components of a multivector $\fieldA$ by $\fieldA$ and $\fieldA'$ in the original and new coordinates respectively.
Along the $i$-th coordinate, the basis element $\ebf_i$ is perturbed to first order by an infinitesimal amount
\begin{equation}
	\ebf_i\times (\deltabf\otimes\xbpert),
\end{equation}
where the Jacobian partial-derivative matrix $\deltabf\otimes\xbpert$ is given by
\begin{equation}
	\deltabf\otimes\xbpert = \sum_{i,j}\Delta_{ii}\partial_i\xpert[j]\wbf_{ij}.
\end{equation}
The $j$-th column of the Jacobian matrix contains the exterior derivative (gradient) of the $j$-th component of the perturbation in the coordinates, $\xpert[j]$. With the identity matrix in \eqref{eq:identity-matrix-m}, which we simply denote as $\mathbf{1}$ since the grade of the vectors in the Jacobian matrix for $\xbpert$ is 1,
and taking into account that $\ebf_i' = \ebf_i + \ebf_i\times (\deltabf\otimes\xbpert)$,  the following relationship between the basis $\{\ebf\}$ and $\{\ebf'\}$ holds for all $i$,
\begin{equation}
	\ebf_i' = \ebf_i\times(\mathbf{1}+\deltabf\otimes\xbpert). \label{eq:trans_cov}
\end{equation}

In the new coordinates, the action in \eqref{eq:action-1a} becomes:
\begin{equation}\label{eq:action-1eb}
  \act_{\fieldA'\cdot\fieldB'} = \int_{\mathcal{R}'}\! \drm^{k+n}\xb'\, (\fieldA'\cdot\fieldB'),
\end{equation}
where $\mathcal{R}'$ is the new integration region and the new differential integration element is 
$\drm^{k+n}\xb'$.

In the following subsections, after expressing the multivectors and the action in the new coordinates, we shall first find the change in action $\delta\act_{\fieldA\cdot\fieldB}$ related to this change of coordinates. Then, we shall link the differential form expressing the change in action with the symmetric, rank-2 stress-energy-momentum tensor. In the process, we will obtain several equivalent, alternative expressions for this tensor and its components, as well as a closed-form formula for its interior derivative (divergence). Finally, we shall relate the vanishing of $\delta\act_{\fieldA\cdot\fieldB}$ induced by the closed nature to the system to the symmetry under the change of coordinates and to a conservation law in terms of the interior derivative of the stress-energy-momentum tensor.

\subsection{Transformation of multivectors of grade $s$}

While the basis vectors transform covariantly according to \eqref{eq:trans_cov}, the multivector field $\fieldA$ transforms contravariantly, that is with the inverse transpose matrix representing the change of coordinates. When $\fieldA$ is a scalar, i.e., with grade 0, it simply holds that 
\begin{equation}\label{eq:29}
	\fieldA' = \fieldA. 
\end{equation}
When $\fieldA$ is a vector field with grade 1, we need two additional facts. First, multiplication by a matrix transpose from the right is equivalent to multiplication by the matrix (untransposed) from the left, as in \eqref{eq:matrix_vector}; second, as the transformation $\xbpert $ is infinitesimal, we have $(\mathbf{1}+\deltabf\otimes\xbpert)^{-1} = \mathbf{1}-\deltabf\otimes\xbpert$. We therefore variously have
\begin{equation}\label{eq:30}
	\fieldA' = \fieldA\times(\mathbf{1} + \deltabf\otimes\xbpert)^{-T} = \fieldA\times(\mathbf{1} - \deltabf\otimes\xbpert)^{T} = (\mathbf{1} - \deltabf\otimes\xbpert)\times\fieldA.
\end{equation}
In passing, and for later use, we note that the derivative operator $\deltabf$ transforms as
\begin{gather}
	\deltabf' =  (\mathbf{1}-\deltabf\otimes\xbpert)\times\deltabf.
\end{gather}

We shall shortly find that an $s$-vector $\fieldA$ transforms in general under this infinitesimal translation as
\begin{equation}\label{eq:fieldA-transf}
	\fieldA' = \fieldA - \trMs\times\fieldA,
\end{equation}
where $\trMs$ is a matrix that varies with the grade $s$, with rows and columns indexed by $s$-tuples, and given by
 \begin{equation}\label{eq:translationMatrixs}
	\trMs = \sum_{I,J\in\mathcal{I}_s} \trcoeff_{I,J}^s\wbf_{I,J}.
 \end{equation}
As in \eqref{eq:29}, we  have $\trMs[0] = 0$ for $s = 0$, as there is only one option for $I = J$, the empty set $\O$ (in other words, $1' = 1$). Similarly, for $s = 1$, the matrix is given by $\trMs[1] = \deltabf\otimes\xbpert$ as in \eqref{eq:30}, and therefore, $\trcoeff_{I,J}^s = \trcoeff_{ij}= \Delta_{ii}\partial_i\xpert[j]$. 

From \eqref{eq:trans_cov}, we find that the basis elements in the new coordinates can be variously written as
\begin{align}
	\ebf_i' 
	&= \ebf_i + \sum_j \partial_i\xpert[j]\ebf_j \label{eq:basis_prime} \\
	&= \sum_j \tau_{i,j}\ebf_j, \quad \text{ with } \tau_{i,j} = \delta_{ij} + \partial_i\xpert[j].
\end{align}
Therefore, the basis element $\ebf_{I}'$ for the $s$-vector in the shifted coordinates, with $I=(i_1,\dotsc,i_s)$, is in turn given by
\begin{align}
	\ebf_{I}' &= \ebf_{i_1}' \wedge \ebf_{i_2}' \dotsm \wedge \ebf_{i_s}' \\
	&= \Biggl(\sum_j \tau_{i_1,j}\ebf_j\Biggr) \wedge \Biggl(\sum_j \tau_{i_2,j}\ebf_j\Biggr) \dotsm \wedge \Biggl(\sum_j \tau_{i_s,j}\ebf_j\Biggr) \label{eq:55} \\
	&= \sum_{J\in\mathcal{I}_s} \det\begin{pmatrix}\tau_{i_1,j_1} & \dotsc & \tau_{i_1,j_s} \\ \vdots & & \vdots \\ \tau_{i_s,j_1} & \dotsc & \tau_{i_s,j_s}\end{pmatrix} \ebf_{J} \label{eq:I-J-det} \\
	&= \sum_{J\in\mathcal{I}_s} \det\tau_{I\otimes J}\, \ebf_{J},\label{eq:eI-eJ-det}
\end{align}
where we have used in \eqref{eq:I-J-det} the relationship between the determinant and the wedge product \cite[Sect.\ 2.2]{flanders1989differentialForms} and in \eqref{eq:I-J-det} we introduced the (linear-algebraic) matrix $\tau_{I\otimes J}$, a submatrix of $\mathbf{1} + \deltabf\otimes\xbpert$, explicitly given by
\begin{equation}
	\tau_{I\otimes J} = \begin{pmatrix}\tau_{i_1,j_1} & \dotsc & \tau_{i_1,j_s} \\ \vdots & & \vdots \\ \tau_{i_s,j_1} & \dotsc & \tau_{i_s,j_s}\end{pmatrix}.
\end{equation}

In the computation of the determinant in \eqref{eq:eI-eJ-det}, we only keep terms up to the first order in the derivatives of $\xbpert$. In order to remove unneeded terms, we first observe that the number of possible overlaps between $I$ and $J$ ranges from 0 to $s$. If there are strictly fewer than $s-1$ overlapping indices, the determinant is zero up to first order, as all summands in the determinant have at least two partial derivatives multiplied together. For each list $I$, we need thus consider only index lists $J$ such that $I = J$ or $\len{I\cap J} = s-1$: 
\begin{align}
	\ebf_{I}' &= \det\tau_{I\otimes I}\, \ebf_{I} + \sum_{J\in\mathcal{I}_s: \len{I\cap J} = s-1} \det\tau_{I\otimes J}\, \ebf_{J}.\label{eq:eI-eJ-det-b}
\end{align}
For $I = J$, the only non zero contribution to the determinant comes from the main diagonal and we directly obtain
\begin{equation}\label{eq:diag-basis_change}
	\det\tau_{I\otimes I} = 1 + \sum_{i\in I} \partial_i\xpert[i].
\end{equation}
As for the case where $|I\cap J| = s-1$, of which there is a total number of $s(k+n-s)$ possibilities, let us define a set $K = I\cap J$ and two indices $i$ and $j$, $i = I\setminus K$ and $j = J\setminus K$, such that $I = \varepsilon(i,K)$ and $J = \varepsilon(j,K)$. 
We can directly infer from \eqref{eq:55} that there is a single non zero contribution to the determinant in this case, namely 
\begin{equation}
	\sigma\bigl(I_{i\leftrightarrow j}\bigr)\,\tau_{i,j}\prod_{k\in K} \tau_{k,k},
\end{equation}
where the signature is that of the permutation that orders the list of indices $I_{i\leftrightarrow j}$ where $i$ in $I$ is replaced by $j$. 
Since $\tau_{i,j} = \partial_i\xpert[j]$ and the terms $\tau_{k,k}$ contribute to the determinant with a 1, we have
\begin{equation}\label{eq:non-diag-basis_change}
	\det\tau_{I\otimes J} = \sigma\bigl(I_{i\leftrightarrow j}\bigr)\,\partial_i\xpert[j].
\end{equation}
As we prove in Appendix \ref{sec:symmetry_ij}, the permutation signature can be expressed various equivalent forms:
\begin{equation}
	\sigma\bigl(I_{i\leftrightarrow j}\bigr) = \sigma\bigl(\varepsilon(i,K)_{i\leftrightarrow j}\bigr) = \sigma\bigl(\varepsilon(j,K)_{j\leftrightarrow i}\bigr) = \sigma\bigl(J_{j\leftrightarrow i}\bigr),
	\label{eq:switch-ij}
\end{equation}
thereby proving that the permutation signature is indeed symmetric in the pair of indices $(i,j)$.
%
%
%
In the same Appendix, we prove the following identities relating the signature in \eqref{eq:switch-ij}, 
\begin{gather}
	\sigma(K,i)\sigma(j,K) = (-1)^{\len{K}}\sigma\bigl(\varepsilon(i,K)_{i\leftrightarrow j}\bigr), \label{eq:98} \\
	\sigma(i,j+K)\sigma(i+K,j) = (-1)^{\len{K}}\sigma\bigl(\varepsilon(i,K)_{i\leftrightarrow j}\bigr). \label{eq:99} 
\end{gather}	

Combining \eqref{eq:diag-basis_change} and \eqref{eq:non-diag-basis_change} into \eqref{eq:eI-eJ-det-b} yields the following expression for the new coordinate basis elements:
\begin{align}
	\ebf_{I}' &= \Biggl(1 + \sum_{i\in I} \partial_i\xpert[i]\Biggr)\, \ebf_{I} + \sum_{K\in\mathcal{I}_{s-1}: i,j\notin K} \sigma\bigl(I_{i\leftrightarrow j}\bigr)\,\partial_i\xpert[j]\, \ebf_{\varepsilon(j,K)}.\label{eq:eI-eJ-det-c}
\end{align}
As we did in \eqref{eq:trans_cov}, we can write \eqref{eq:eI-eJ-det-c} as 
\begin{equation}
	\ebf_I' = \ebf_I\times(\mathbf{1}_s+\trMs), \label{eq:trans_cov_s}
\end{equation}
where $\mathbf{1}_s$ is the identity matrix for grade-$s$ multivectors, $\mathbf{1}_s = \sum_{I}\Delta_{II}\wbf_{I,I}$, and 
the matrix $\trMs$ is such that 
\begin{equation}
	\ebf_I\times \trMs = \Biggl(\sum_{i\in I} \partial_i\xpert[i]\Biggr)\, \ebf_{I} + \sum_{K\in\mathcal{I}_{s-1}: i,j\notin K} \sigma\bigl(I_{i\leftrightarrow j}\bigr)\,\partial_i\xpert[j]\, \ebf_{\varepsilon(j,K)}.
\end{equation}
Taking into account the definition of the product $\times$, the matrix $\trMs$ is therefore given by
\begin{align}
	\trMs &= \Delta_{II}\ebf_{I}\otimes \Biggl(\sum_{i\in I} \partial_i\xpert[i]\Biggr)\, \ebf_{I} + \Delta_{II}\ebf_{I}\otimes \sum_{K\in\mathcal{I}_{s-1}: i,j\notin K} \sigma\bigl(I_{i\leftrightarrow j}\bigr)\,\partial_i\xpert[j]\, \ebf_{\varepsilon(j,K)} \\
	&= \Delta_{II}\Biggl(\sum_{i\in I} \partial_i\xpert[i]\Biggr)\, \wbf_{I,I} + \Delta_{II}\sum_{K\in\mathcal{I}_{s-1}: i,j\notin K} \sigma\bigl(I_{i\leftrightarrow j}\bigr)\,\partial_i\xpert[j]\, \wbf_{I,\varepsilon(j,K)}.
\end{align}
For later use, it will prove convenient to express the matrix $\trMs$ in the equivalent form:
\begin{equation}
	\trMs = \sum_i\Delta_{ii}\partial_i\xpert[i]\trMs(i,i) + \sum_{i,j:i\neq j}\Delta_{ii}\partial_i\xpert[j]\trMs(i,j), \label{eq:g_eps_s}
\end{equation}
where the matrices $\trMs(i,i)$ and $\trMs(i,j)$, for all $i$ and $j$, with $j\neq i$, are, respectively, given by
\begin{gather}
	\trMs(i,i) = \sum_{K\in{\mathcal{I}_{s-1}}:i\notin K}\Delta_{KK}\wbf_{\varepsilon(i,K),\varepsilon(i,K)} \label{eq:g_eps_s_ii} \\ 
	\trMs(i,j) = \sum_{K\in{\mathcal{I}_{s-1}}:i,j\notin K}\Delta_{KK}\sigma\bigl(\varepsilon(i,K)_{i\leftrightarrow j}\bigr)\wbf_{\varepsilon(i,K),\varepsilon(j,K)}. \label{eq:g_eps_s_ij}
\end{gather}

\subsection{Transformation of the action: stress-energy-momentum tensor}

We now return to the expression of the action in the new coordinates up to derivatives of first order in $\xbpert$. We first note that the differential $\drm^{k+n}\xb$ appearing in the action is expressed in the new coordinates using \eqref{eq:diag-basis_change} with $I$ including all space-time indices, that is $\len{I} = k+n$, so that we directly obtain
\begin{equation}\label{eq:drimprime}
	\drm^{k+n}\xb' = \drm^{k+n}\xb\bigl(1+\deltabf\cdot\xbpert).
\end{equation}
In the new coordinates, the integration region is denoted as $\mathcal{R}'$; for an infinitesimal translation, the regions $\mathcal{R}$ and $\mathcal{R}'$ differ only on the boundary of the former, which is located far from the origin.
With the identity in \eqref{eq:drimprime}, together with \eqref{eq:fieldA-transf} for $\fieldA$ and $\fieldB$, and neglecting higher-order derivative terms, 
Eq.\ \eqref{eq:action-1eb} becomes
\begin{align}\label{eq:action-2e}
  \act_{\fieldA'\cdot\fieldB'} &= \int_{\mathcal{R}'}\! \drm^{k+n}\xb'\, (\fieldA'\cdot\fieldB') \\
  &= \int_{\mathcal{R}}\! \drm^{k+n}\xb\bigl(1+\deltabf\cdot\xbpert)\, \Bigl(\bigl((\mathbf{1}-\trMs)\times\fieldA\bigr)\cdot\bigl((\mathbf{1}-\trMs)\times\fieldB\bigr)\Bigr) \label{eq:70} \\
  &= \int_{\mathcal{R}}\! \drm^{k+n}\xb\, (\fieldA\cdot\fieldB) + \int_{\mathcal{R}}\! \drm^{k+n}\xb(\deltabf\cdot\xbpert)\,(\fieldA\cdot\fieldB) - \int_{\mathcal{R}}\! \drm^{k+n}\xb\, \bigl(\trMs\times\fieldA)\cdot\fieldB\bigr) - \int_{\mathcal{R}}\! \drm^{k+n}\xb\, \bigl(\fieldA\cdot(\trMs\times\fieldB)\bigr) \\
  &= \act_{\fieldA\cdot\fieldB} + \delta \act_{\fieldA\cdot\fieldB},
\end{align}
where in \eqref{eq:70} we used that the difference in integration regions $\mathcal{R}'$ and $\mathcal{R}$ lies far from the origin, coupled with the rapid decay of the fields, to replace $\mathcal{R}'$ by $\mathcal{R}$, and finally defined the change in action $\delta \act_{\fieldA\cdot\fieldB}$ is given by
\begin{equation}\label{eq:delta_action-3e}
  \delta \act_{\fieldA\cdot\fieldB} = \int_{\mathcal{R}}\! \drm^{k+n}\xb(\deltabf\cdot\xbpert)\,(\fieldA\cdot\fieldB) - \int_{\mathcal{R}}\! \drm^{k+n}\xb\, \bigl((\trMs\times\fieldA)\cdot\fieldB\bigr) - \int_{\mathcal{R}}\! \drm^{k+n}\xb\, \bigl(\fieldA\cdot(\trMs\times\fieldB)\bigr).
\end{equation}

We next prove the existence of a symmetric, rank-2 tensor $\abset$ such that
\begin{equation}\label{eq:delta_action-4e}
	\delta \act_{\fieldA\cdot\fieldB} = \int_\mathcal{R}\! \drm^{k+n}\xb\,(\deltabf\otimes\xbpert)\cdot\abset.
\end{equation}
The tensor $\abset$ may in fact be identified with the symmetric stress-energy momentum tensor associated with the Lagrangian density $\fieldA\cdot\fieldB$. In terms of the components of $\abset$, we have
\begin{equation}\label{eq:abset-uij}
	\abset = \sum_{i\leq j} \abseti_{ij}\ubf_{ij},
\end{equation}
where on- and off-diagonal components, $\abseti_{ii}$ and $\abseti_{ij}$, are respectively given by
\begin{gather}
	\abseti_{ii} = \Delta_{ii}\Biggl(\sum_{K\in{\mathcal{I}_{s}}:i\notin K}\Delta_{KK}\fieldAi_{K}\fieldBi_{K}-\sum_{K\in{\mathcal{I}_{s}}:i\in K}\Delta_{KK}\fieldAi_{K}\fieldBi_{K}\Biggr), \label{eq:abset_ii} \\
	\abseti_{ij} = -\sum_{K\in{\mathcal{I}_{s-1}}:i,j\notin K}\Delta_{KK}\sigma\bigl(\varepsilon(i,K)_{i\leftrightarrow j}\bigr)\bigl(\fieldAi_{\varepsilon(i,K)}\fieldBi_{\varepsilon(j,K)}+\fieldBi_{\varepsilon(i,K)}\fieldAi_{\varepsilon(j,K)}\bigr). \label{eq:abset_ij}
\end{gather}
As befits a symmetric rank-2 tensor, the coefficient $\abseti_{ij}$ in \eqref{eq:abset_ij} does not change under permutation of $i$ and $j$:
the factor $\bigl(\fieldAi_{\varepsilon(i,K)}\fieldBi_{\varepsilon(j,K)}+\fieldBi_{\varepsilon(i,K)}\fieldAi_{\varepsilon(j,K)}\bigr)$  is clearly symmetric and the permutation signature $\sigma\bigl(\varepsilon(i,K)_{i\leftrightarrow j}\bigr)$ is also symmetric in $i,j$, as shown in \eqref{eq:switch-ij}.

In order to prove \eqref{eq:delta_action-4e}, we shall establish the following identity of differential forms appearing in the integrand,
\begin{equation}\label{eq:diff-forms-Tab}
	(\deltabf\otimes\xbpert)\cdot\abset = (\deltabf\cdot\xbpert)\,(\fieldA\cdot\fieldB) -  (\trMs\times\fieldA)\cdot\fieldB - \fieldA\cdot(\trMs\times\fieldB).
\end{equation}
Expanding the left-hand side of \eqref{eq:diff-forms-Tab} and using the dot product formula for $\wbf_{ij}$ and $\ubf_{ij}$ in \eqref{eq:dot_tensor_symm} yields
\begin{align}
	(\deltabf\otimes\xbpert)\cdot\abset &= \Biggl(\sum_{i,j}\Delta_{ii}\partial_{i}\xpert[j]\wbf_{ij}\Biggr)\cdot\Biggl(\sum_{i'\leq j'}\abseti_{i'j'}\ubf_{i'j'} \Biggr) \\
	&= \sum_{i,j}\Delta_{ii}\partial_{i}\xpert[j]\Delta_{ii}\Delta_{jj}\abseti_{\varepsilon(i,j)}.\label{eq:diff-forms-Tab-lhs}
\end{align}
At this point, if we expand the right-hand side of \eqref{eq:diff-forms-Tab} into a summation whose terms are indexed by the pair $(i,j)$  such that each of these summands has the form of a coefficient multiplying $\Delta_{ii}\partial_{i}\xpert[j]$, we can directly read $\abseti_{\varepsilon(i,j)}$ from this multiplicative coefficient. More precisely, using the definition of $\deltabf\cdot\xbpert$ together with \eqref{eq:g_eps_s} yields
\begin{align}
	\Biggl(\sum_i\partial_i\xpert[i]\Biggr)\,(\fieldA\cdot\fieldB) -  &\Biggl(\sum_i\Delta_{ii}\partial_i\xpert[i]\trMs(i,i)\times\fieldA + \sum_{i,j:i\neq j}\Delta_{ii}\partial_i\xpert[j]\trMs(i,j)\times\fieldA\Biggr)\cdot\fieldB - \notag\\ & \fieldA\cdot\Biggl(\sum_i\Delta_{ii}\partial_i\xpert[i]\trMs(i,i)\times\fieldB + \sum_{i,j:i\neq j}\Delta_{ii}\partial_i\xpert[j]\trMs(i,j)\times\fieldB\Biggr).\label{eq:diff-forms-Tab-rhs}
\end{align}

It will prove convenient to distinguish the cases $i = j$ and $i \neq j$. First, for $i = j$ and writing the relationship between the diagonal terms in \eqref{eq:diff-forms-Tab-lhs} and \eqref{eq:diff-forms-Tab-rhs}, we obtain
\begin{align}
	\Delta_{ii}\partial_{i}\xpert[i]\Delta_{ii}\Delta_{ii}\abseti_{ii} &= \partial_i\xpert[i]\,(\fieldA\cdot\fieldB) - \bigl(\Delta_{ii}\partial_i\xpert[i]\trMs(i,i)\times\fieldA\bigr)\cdot\fieldB - \bigl(\Delta_{ii}\partial_i\xpert[i]\trMs(i,i)\times\fieldB\bigr)\cdot\fieldA \\
	\abseti_{ii} &= \Delta_{ii}(\fieldA\cdot\fieldB) - \bigl(\trMs(i,i)\times\fieldA\bigr)\cdot\fieldB - \bigl(\trMs(i,i)\times\fieldB\bigr)\cdot\fieldA. \label{eq:Tab_ii_proof}
\end{align}

Taking into account the definition of $\trMs[s](i,i)$ in \eqref{eq:g_eps_s_ii} and carrying out the matrix products in \eqref{eq:Tab_ii_proof}, we obtain
\begin{align}
	\fieldA\cdot\bigl(\trMs[s](i,i)\times \fieldB\bigr) &= \fieldA\cdot\Biggl(\sum_{K\in{\mathcal{I}_{s-1}}:i\notin K}\Delta_{ii}\ebf_{\varepsilon(i,K)}\fieldBi_{\varepsilon(i,K)}\Biggr) \\
	&= \sum_{K\in{\mathcal{I}_{s-1}}:i\notin K}\Delta_{KK}\fieldAi_{\varepsilon(i,K)}\fieldBi_{\varepsilon(i,K)}, \label{eq:45} \\
	\fieldB\cdot\bigl(\trMs[s](i,i)\times \fieldA\bigr) &= \sum_{K\in{\mathcal{I}_{s-1}}:i\notin K}\Delta_{KK}\fieldBi_{\varepsilon(i,K)}\fieldAi_{\varepsilon(i,K)}.\label{eq:46}
\end{align}
Using the definition of the dot product together with \eqref{eq:45}--\eqref{eq:46} back in \eqref{eq:Tab_ii_proof} yields
\begin{align}
	\abseti_{ii} &= \Delta_{ii}\sum_{K\in{\mathcal{I}_{s}}}\Delta_{KK}\fieldBi_{K}\fieldAi_{K}-2\sum_{K\in{\mathcal{I}_{s-1}}:i\notin K}\Delta_{KK}\fieldBi_{\varepsilon(i,K)}\fieldAi_{\varepsilon(i,K)} \\
	&= \Delta_{ii}\Biggl(\sum_{K\in{\mathcal{I}_{s}}}\Delta_{KK}\fieldBi_{K}\fieldAi_{K}-\sum_{K\in{\mathcal{I}_{s}}:i\in K}\Delta_{KK}\fieldBi_{K}\fieldAi_{K}\Biggr) \\
&= \Delta_{ii}\Biggl(\sum_{K\in{\mathcal{I}_{s}}:i\notin K}\Delta_{KK}\fieldBi_{K}\fieldAi_{K}-\sum_{K\in{\mathcal{I}_{s}}:i\in K}\Delta_{KK}\fieldBi_{K}\fieldAi_{K}\Biggr),
\end{align}
namely the desired expression for $\abseti_{ii}$ in \eqref{eq:abset_ii}.

In an analogous manner, we write the relationship between the off-diagonal terms in \eqref{eq:diff-forms-Tab-lhs} and \eqref{eq:diff-forms-Tab-rhs}, $i\neq j$, to obtain the following expressions for the off-diagonal tensor elements:
\begin{align}
	\Delta_{ii}\partial_{i}\xpert[j]\Delta_{ii}\Delta_{jj}\abseti_{\varepsilon(i,j)} &= - \bigl(\Delta_{ii}\partial_i\xpert[j]\trMs(i,j)\times\fieldA\bigr)\cdot\fieldB - \bigl(\Delta_{ii}\partial_i\xpert[j]\trMs(i,j)\times\fieldB\bigr)\cdot\fieldA \\
	\abseti_{\varepsilon(i,j)} &= - \Delta_{ii}\Delta_{jj}\bigl(\trMs(i,j)\times\fieldA\bigr)\cdot\fieldB - \Delta_{ii}\Delta_{jj}\bigl(\trMs(i,j)\times\fieldB\bigr)\cdot\fieldA, \label{eq:Tab_ij_proof}
\end{align}
where one should verify that the right-hand side of \eqref{eq:Tab_ij_proof} is indeed symmetric in $i$ and $j$, as we will shortly do.
As before, using the definition of $\trMs[s](i,j)$ in \eqref{eq:g_eps_s_ij} and carrying out the matrix products in \eqref{eq:Tab_ij_proof}, we obtain
\begin{align}
	\fieldA\cdot\bigl(\trMs[s](i,j)\times \fieldB\bigr) &= \fieldA\cdot\Biggl(\sum_{K\in{\mathcal{I}_{s-1}}:i,j\notin K}\Delta_{jj}\sigma\bigl(\varepsilon(i,K)_{i\leftrightarrow j}\bigr)\ebf_{\varepsilon(i,K)}\fieldBi_{\varepsilon(j,K)}\Biggr) \\
 	&= \sum_{K\in{\mathcal{I}_{s-1}}:i,j\notin K}\Delta_{jj}\Delta_{ii}\Delta_{KK}\sigma\bigl(\varepsilon(i,K)_{i\leftrightarrow j}\bigr)\fieldAi_{\varepsilon(i,K)}\fieldBi_{\varepsilon(j,K)}, \label{eq:51} \\
	\fieldB\cdot\bigl(\trMs[s](i,j)\times \fieldA\bigr) &= \sum_{K\in{\mathcal{I}_{s-1}}:i,j\notin K}\Delta_{jj}\Delta_{ii}\Delta_{KK}\sigma\bigl(\varepsilon(i,K)_{i\leftrightarrow j}\bigr)\fieldBi_{\varepsilon(i,K)}\fieldAi_{\varepsilon(j,K)}. \label{eq:52}
\end{align}
Again, replacing both \eqref{eq:51} and \eqref{eq:52} back in \eqref{eq:Tab_ij_proof} leads to the desired expression for $\abseti_{ij}$, namely \eqref{eq:abset_ij}:
\begin{equation}
	\abseti_{ij} = -\sum_{K\in{\mathcal{I}_{s-1}}:i,j\notin K}\Delta_{KK}\sigma\bigl(\varepsilon(i,K)_{i\leftrightarrow j}\bigr)\bigl(\fieldAi_{\varepsilon(i,K)}\fieldBi_{\varepsilon(j,K)}+\fieldBi_{\varepsilon(i,K)}\fieldAi_{\varepsilon(j,K)}\bigr).
\end{equation}
In this equation, the factor $\bigl(\fieldAi_{\varepsilon(i,K)}\fieldBi_{\varepsilon(j,K)}+\fieldBi_{\varepsilon(i,K)}\fieldAi_{\varepsilon(j,K)}\bigr)$ clearly remains unchanged after permutation of $i$ and $j$. As for $\sigma\bigl(\varepsilon(i,K)_{i\leftrightarrow j}\bigr)$, we proved in \eqref{eq:switch-ij} that the permutation signature is indeed symmetric in $i,j$. 
Thus, the coefficient $\abseti_{ij}$ does not change under permutation of $i$ and $j$, as befits a symmetric rank-2 tensor.	
	
\subsection{An alternative, coordinate-free, expression for the tensor}

In this section, we provide an alternative, coordinate-free, expression for the symmetric stress-energy-momentum tensor $\abset$. This expression given in terms of the operations $\odot$ and $\owedge$ is defined in \eqref{eq:def-odot}--\eqref{eq:def-owedge}, as follows:
\begin{equation}\label{eq:abset-oo}
	\abset = (-1)^{s}\bigl(\fieldA\odot\fieldB + \fieldA\owedge\fieldB\bigr).  
\end{equation}
The discussion after the definitions of \eqref{eq:def-owedge} and \eqref{eq:def-odot} proves that the tensor field in \eqref{eq:abset-oo} is indeed symmetric.

In order to prove the expression in \eqref{eq:abset-oo}, we make use of the tensor components in \eqref{eq:abset_ii}--\eqref{eq:abset_ij}. 
From the definitions in \eqref{eq:def-owedge} and \eqref{eq:def-odot}, we may, respectively, write the coefficients multiplying the basis element $\ubf_{ij}$ as:
\begin{align}
	(\Delta_{ii}\ebf_i\wedge\fieldA)\cdot(\fieldB\wedge\ebf_j\Delta_{jj}) &= \Biggl(\sum_{I\in\mathcal{I}_{s}}\Delta_{ii}\fieldAi_I\ebf_i\wedge\ebf_I\Biggr)\cdot\Biggl(\sum_{I\in\mathcal{I}_{s}}\Delta_{jj}\fieldBi_I\ebf_I\wedge\ebf_j\Biggr)	\\ 
	&= \Biggl(\sum_{I\in\mathcal{I}_{s}:i\notin I}\Delta_{ii}\sigma(i,I)\fieldAi_I\ebf_{\varepsilon(i,I)}\Biggr)\cdot\Biggl(\sum_{I\in\mathcal{I}_{s}:j\notin I}\Delta_{jj}\sigma(I,j)\fieldBi_I\ebf_{\varepsilon(j,I)}\Biggr), \label{eq:62}
\end{align}
where we carried out the exterior products in each of the factors, and 
\begin{align}
	(\Delta_{ii}\ebf_i\lintprod\fieldA)\cdot(\fieldB\rintprod\ebf_j\Delta_{jj}) &= \Biggl(\sum_{I\in\mathcal{I}_{s}}\Delta_{ii}\fieldAi_I\ebf_i\lintprod\ebf_I\Biggr)\cdot\Biggl(\sum_{I\in\mathcal{I}_{s}}\Delta_{jj}\fieldBi_I\ebf_I\rintprod\ebf_j\Biggr)	\\ 
	&= \Biggl(\sum_{I\in\mathcal{I}_{s}:i\in I}\sigma(I\setminus i,i)\fieldAi_I\ebf_{I\setminus i}\Biggr)\cdot\Biggl(\sum_{I\in\mathcal{I}_{s}:j\in I}\sigma(j,I\setminus j)\fieldBi_I\ebf_{I\setminus j}\Biggr), \label{eq:60} 
\end{align}
where we similarly carried out the interior products in each of the factors.

In order to evaluate the dot products in \eqref{eq:62} and \eqref{eq:60}, it proves convenient to distinguish between the on- and off-diagonal components. First, for $i = j$ and \eqref{eq:62}, we obtain
\begin{align}
	(\Delta_{ii}\ebf_i\wedge\fieldA)\cdot(\fieldB\wedge\ebf_i\Delta_{ii}) &= \Biggl(\sum_{I\in\mathcal{I}_{s}:i\notin I}\Delta_{ii}\sigma(i,I)\fieldAi_I\ebf_{\varepsilon(i,I)}\Biggr)\cdot\Biggl(\sum_{I\in\mathcal{I}_{s}:i\notin I}\Delta_{ii}\sigma(I,i)\fieldBi_I\ebf_{\varepsilon(i,I)}\Biggr) \\
	&= \sum_{I\in\mathcal{I}_{s}:i\notin I}\Delta_{ii}\Delta_{II}\sigma(I,i)\sigma(i,I)\fieldAi_I\fieldBi_I \\
	&= (-1)^{s}\sum_{I\in\mathcal{I}_{s}:i\notin I}\Delta_{ii}\Delta_{II}\fieldAi_I\fieldBi_I, \label{eq:68}
\end{align}
where we used the identity $\sigma(i,I)\sigma(I,i) = (-1)^{\len{I}} = (-1)^{s}$, $i\notin I$. Similarly, for $i = j$ and \eqref{eq:60}, we obtain
\begin{align}
	(\Delta_{ii}\ebf_i\lintprod\fieldA)\cdot(\fieldB\rintprod\ebf_i\Delta_{ii}) &= \Biggl(\sum_{I\in\mathcal{I}_{s}:i\in I}\sigma(I\setminus i,i)\fieldAi_I\ebf_{I\setminus i}\Biggr)\cdot\Biggl(\sum_{I\in\mathcal{I}_{s}:j\in I}\sigma(j,I\setminus j)\fieldBi_I\ebf_{I\setminus j}\Biggr) \\
	&= \sum_{I\in\mathcal{I}_{s}:i\in I}\Delta_{ii}\Delta_{II}\sigma(i,I\setminus i)\sigma(I\setminus i,i)\fieldAi_I\fieldBi_I \\
	&= (-1)^{s-1}\sum_{I\in\mathcal{I}_{s}:i\in I}\Delta_{ii}\Delta_{II}\fieldAi_I\fieldBi_I, \label{eq:65}
\end{align}
where we again used the identity $\sigma(i,K)\sigma(K,i) = (-1)^{s-1}$, for $i\notin K$. 
Combining  \eqref{eq:68} with \eqref{eq:65} and \eqref{eq:abset_ii} gives:
\begin{align}
	(\Delta_{ii}\ebf_i\wedge\fieldA)\cdot(\fieldB\wedge\ebf_i\Delta_{ii}) + (\Delta_{ii}\ebf_i\lintprod\fieldA)\cdot(\fieldB\rintprod\ebf_i\Delta_{ii})  &= (-1)^{s-1}\Delta_{ii}\Biggl(\sum_{I\in\mathcal{I}_{s}:i\in I}\Delta_{II}\fieldAi_I\fieldBi_I - \sum_{I\in\mathcal{I}_{s}:i\notin I}\Delta_{II}\fieldAi_I\fieldBi_I\Biggr) \\
	&= (-1)^{s}\abseti_{ii},
\end{align}
Noticing that this expression is actually symmetric in $\fieldA$ and $\fieldB$, its average with itself with the roles of $\fieldA$ and $\fieldB$ interchanged gives the same expression, in correspondence with \eqref{eq:abset-oo}, as far as the on-diagonal terms are concerned.

Concerning the off-diagonal components, $i \neq j$, we start by evaluating \eqref{eq:60} to directly obtain:
\begin{align}
	(\Delta_{ii}\ebf_i\lintprod\fieldA)\cdot(\fieldB\rintprod\ebf_j\Delta_{jj}) &= \sum_{K\in\mathcal{I}_{s-1}:i,j\notin K}\Delta_{KK}\sigma(K,i)\sigma(j,K)\fieldAi_{\varepsilon(i,K)}\fieldBi_{\varepsilon(j,K)}. \label{eq:71} 
\end{align}
Similarly, we evaluate \eqref{eq:62} and expand the dot product therein to obtain:
\begin{align}
	(\Delta_{ii}\ebf_i\wedge\fieldA)\cdot(\fieldB\wedge\ebf_j\Delta_{jj}) &= \Biggl(\sum_{I\in\mathcal{I}_{s}:i\notin I}\Delta_{ii}\sigma(i,I)\fieldAi_I\ebf_{\varepsilon(i,I)}\Biggr)\cdot\Biggl(\sum_{I\in\mathcal{I}_{s}:j\notin I}\Delta_{jj}\sigma(I,j)\fieldBi_I\ebf_{\varepsilon(j,I)}\Biggr) \\
	 &= \sum_{K\in\mathcal{I}_{s-1}:i,j\notin K}\Delta_{KK}\sigma(i,j+K)\sigma(i+K,j)\fieldAi_{j+K}\fieldBi_{i+K}.	\label{eq:73}
\end{align}
Using \eqref{eq:98} and \eqref{eq:99}, we combine \eqref{eq:71} together with \eqref{eq:73} and \eqref{eq:abset_ij} to write
\begin{align}
	(\Delta_{ii}\ebf_i\lintprod\fieldA)\cdot(\fieldB\rintprod\ebf_j\Delta_{jj}) &+ (\Delta_{ii}\ebf_i\wedge\fieldA)\cdot(\fieldB\wedge\ebf_j\Delta_{jj}) 
	\nonumber \\&
	= (-1)^{s-1}\mkern-24mu\sum_{K\in\mathcal{I}_{s-1}:i,j\notin K}\mkern-18mu\Delta_{KK}\sigma\bigl(\varepsilon(i,K)_{i\leftrightarrow j}\bigr)(\fieldAi_{i+K}\fieldBi_{j+K} + \fieldAi_{j+K}\fieldBi_{i+K}) \\
	&= (-1)^{s}\abseti_{ij},
\end{align}
once again in correspondence with the off-diagonal terms in \eqref{eq:abset-oo}.

\subsection{Conservation law for energy-momentum}

The next step is to apply a generalized Leibniz rule to rewrite the integrand in \eqref{eq:delta_action-4e} in convenient way. 
For a vector $\xbpert$ of grade 1 and a symmetric tensor of rank 2, denoted by $\Tb$, the following generalized Leibniz rule holds:
\begin{equation}
	\deltabf\cdot(\xbpert\lintprod\Tb) = \xbpert\cdot(\deltabf\lintprod\Tb)+ (\deltabf\otimes\xbpert)\cdot\Tb,
	\label{eq:67}
\end{equation}
where the interior product between a vector and a symmetric tensor is computed according to \eqref{eq:left-int-prod-multiv-tens}.
This equation relates scalars, or zero-grade multivectors, on both sides. It is proved in Appendix \ref{sec:leibniz-mixed}.
Using \eqref{eq:67} and the vanishing at infinity of $\xbpert$ to directly neglect the term in the left-hand side of \eqref{eq:67}, we express $\delta\act_{\fieldA\cdot\fieldB}$ in terms of the interior derivative (or divergence) of the symmetric stress-energy momentum tensor $\abset$ as
\begin{align}
	\delta \act_{\fieldA\cdot\fieldB} &= -\int_\mathcal{R}\! \drm^{k+n}\xb\,(\deltabf\lintprod\abset)\cdot\xbpert.  
\end{align}
Assuming that infinitesimal space-time translations are a symmetry of the whole system and that the fields decay sufficiently fast at the boundary of the region $\mathcal{R}$, the fact that the variation of the action $\delta\semt{sys}$ must be zero for all infinitesimal perturbations $\xbpert$ implies that the interior derivative of the overall tensor $\semt{sys}$ in \eqref{eq:sem_total} is zero, 
\begin{equation}
	\deltabf\lintprod\semt{sys} = 0,
\end{equation}
which yields a conservation law for the energy-momentum of the system under consideration.


Using the definition of interior product in \eqref{eq:left-int-prod-multiv-tens}, we directly obtain:
\begin{align}
	\deltabf\lintprod\abset &= \sum_{i,j} \partial_i\abseti_{\varepsilon(i,j)}\ebf_j \\
	&= \sum_i \Biggl(\partial_i\abseti_{i,i}\ebf_i + \sum_{j\neq i} \partial_i\abseti_{\varepsilon(i,j)}\ebf_j\Biggr), \label{eq:int-der-abset}
\end{align}
where we may use the formulas for the tensor components \eqref{eq:abset_ii} and \eqref{eq:abset_ij} to write:
\begin{align}
	\partial_i\abseti_{i,i}& = \Delta_{ii}\Biggl(\sum_{K\in{\mathcal{I}_{s}}:i\notin K}\Delta_{KK}\partial_i(\fieldBi_{K}\fieldAi_{K})-\sum_{K\in{\mathcal{I}_{s}}:i\in K}\Delta_{KK}\partial_i(\fieldBi_{K}\fieldAi_{K})\Biggr), \label{eq:int-der-abset-ii} \\
	\partial_i\abseti_{{\varepsilon(i,j)}}& = -\sum_{K\in{\mathcal{I}_{s-1}}:i,j\notin K}\Delta_{KK}\sigma\bigl(\varepsilon(i,K)_{i\leftrightarrow j}\bigr)\partial_i\bigl(\fieldAi_{\varepsilon(i,K)}\fieldBi_{\varepsilon(j,K)}+\fieldBi_{\varepsilon(i,K)}\fieldAi_{\varepsilon(j,K)}\bigr). \label{eq:int-der-abset-ij}
\end{align}

In addition to the identities in \eqref{eq:int-der-abset}--\eqref{eq:int-der-abset-ij}, we provide in this section a simple, coordinate-free, closed-form expression for the interior derivative of the stress-energy-momentum tensor $\abset$:
\begin{equation}
	\deltabf\lintprod\abset 
	= \fieldA\lintprod(\deltabf\wedge\fieldB) + \fieldB\lintprod(\deltabf\wedge\fieldA) -\fieldA\rintprod(\deltabf\lintprod\fieldB)  - \fieldB\rintprod (\deltabf\lintprod\fieldA). \label{eq:int-der-abset-cf}
\end{equation}

As first step in our proof of \eqref{eq:int-der-abset-cf}, we use \eqref{eq:abset-oo} in order to rewrite the interior derivative of $\abset$,
\begin{equation}
	\deltabf\lintprod\abset = (-1)^{s}\bigl(\deltabf\lintprod(\fieldA\owedge\fieldB) + \deltabf\lintprod(\fieldA\odot\fieldB) \bigr), \label{eq:114}
\end{equation}
and analyze the interior derivatives of $\fieldA\owedge\fieldB$ and $\fieldA\odot\fieldB$. Putting the definition of $\owedge$ in \eqref{eq:def-owedge} into \eqref{eq:int-der-abset}, we get:
\begin{align}
	\deltabf\lintprod(\fieldA\owedge\fieldB) 
	&= \sum_{i,j} \Delta_{ii}\Delta_{jj}\partial_i\bigl((\ebf_{\min(i,j)}\wedge\fieldA)\cdot(\fieldB\wedge\ebf_{\max(i,j)})\bigr)\ebf_j \\
	&= \sum_{i}\Biggl(\partial_i\bigl((\ebf_{i}\wedge\fieldA)\cdot(\fieldB\wedge\ebf_{i})\bigr)\ebf_i\Biggr) + \sum_i\sum_{j\neq i}\Biggl(\Delta_{ii}\Delta_{jj}\partial_i\bigl((\ebf_{\min(i,j)}\wedge\fieldA)\cdot(\fieldB\wedge\ebf_{\max(i,j)})\bigr)\ebf_j\Biggr), \label{eq:116}
\end{align}
where we wrote the equations in a convenient way as a double summation over the pair of indices $i$ and $j$.

We now use \eqref{eq:68} to evaluate the first summation over $i$ in \eqref{eq:116},
\begin{align}
	\sum_{i} \partial_i\bigl((\ebf_{i}\wedge\fieldA)\cdot(\fieldB\wedge\ebf_{i})\bigr)\ebf_i &= (-1)^{s}\sum_{i} \sum_{I\in\mathcal{I}_{s}:i\notin I}\Delta_{ii}\Delta_{II}\partial_i\bigl(\fieldAi_I\fieldBi_I\bigr)\ebf_i. \label{eq:117}
\end{align}
In a similar manner, we use \eqref{eq:73} to evaluate the second summation, over $i$ and $j\neq i$, in \eqref{eq:116},
\begin{align}
	\sum_{i}\sum_{j\neq i} \Delta_{ii}\Delta_{jj}\partial_i&\bigl((\ebf_{\min(i,j)}\wedge\fieldA)\cdot(\fieldB\wedge\ebf_{\max(i,j)})\bigr)\ebf_j = \notag \\
	&= \sum_{i}\sum_{j\neq i}\sum_{K\in\mathcal{I}_{s-1}:i,j\notin K}\Delta_{KK}\sigma\bigl(K,\min(i,j)\bigr)\sigma\bigl(\max(i,j),K\bigr)\partial_i\bigl(\fieldAi_{\varepsilon(\max(i,j),K)}\fieldBi_{\varepsilon(\min(i,j),K)}\bigr)\ebf_j \\
	&= (-1)^{s-1}\sum_{i}\sum_{j\neq i}\sum_{K\in\mathcal{I}_{s-1}:i,j\notin K}\Delta_{KK}\sigma(i,K)\sigma(j,K)\partial_i\bigl(\fieldAi_{\varepsilon(\max(i,j),K)}\fieldBi_{\varepsilon(\min(i,j),K)}\bigr)\ebf_j, \label{eq:119}
\end{align}
where we used that $\sigma(K,\ell) = (-1)^{\len{K}}\sigma(\ell,K)$, and that $\sigma\bigl(\min(i,j),K\bigr)\sigma\bigl(\max(i,j),K\bigr) = \sigma(i,K)\sigma(j,K)$ as $i\neq j$.

We now relate \eqref{eq:116} to a closed-form, coordinate-free expression. First, let us evaluate $\fieldA\lintprod(\deltabf\wedge\fieldB)$, that is,
\begin{align}
	\fieldA\lintprod(\deltabf\wedge\fieldB) &= \Biggl(\sum_{J\in\mathcal{I}_{s}}\fieldAi_J\ebf_J\Biggr)\lintprod\Biggl(\sum_{i,I\in\mathcal{I}_{s}:i\notin I}\Delta_{ii}\sigma(i,I)\partial_i \fieldBi_I\ebf_{\varepsilon(i,I)}\Biggr) \\
	&= \sum_{i,I\in\mathcal{I}_{s}:i\notin I}\sum_{J\in\mathcal{I}_{s}}\Delta_{ii}\sigma(i,I)\fieldAi_J\partial_i \fieldBi_I\ebf_J\lintprod\ebf_{\varepsilon(i,I)} \\
	&= \sum_{i}\sum_{I,J\in\mathcal{I}_{s}:i\notin I,i\in J}\Delta_{ii}\sigma(i,I)\fieldAi_J\partial_i \fieldBi_I\ebf_J\lintprod\ebf_{\varepsilon(i,I)} + \sum_{i}\sum_{I,J\in\mathcal{I}_{s}:i\notin I,i\notin J}\Delta_{ii}\sigma(i,I)\fieldAi_J\partial_i \fieldBi_I\ebf_J\lintprod\ebf_{\varepsilon(i,I)}, \label{eq:122}
\end{align}
where we split the summation over $J$ in two parts, depending on whether $i \in J$ or not. Let us first focus and evaluate the second summation over $i$ in \eqref{eq:122}, namely
\begin{align}
	\sum_{i}\sum_{I,J\in\mathcal{I}_{s}:i\notin I,i\notin J}\Delta_{ii}\sigma(i,I)\fieldAi_J\partial_i \fieldBi_I\ebf_J\lintprod\ebf_{\varepsilon(i,I)} &=
 \sum_{i}\sum_{I\in\mathcal{I}_{s}:i\notin I}\Delta_{ii}\sigma(i,I)\fieldAi_I\partial_i \fieldBi_I\ebf_I\lintprod\ebf_{\varepsilon(i,I)} \label{eq:123} \\
	&= \sum_{i}\sum_{I\in\mathcal{I}_{s}:i\notin I}\Delta_{ii}\Delta_{II}\fieldAi_I\partial_i \fieldBi_I\ebf_i, \label{eq:124}
\end{align}
where we used in \eqref{eq:123} that the constraints that $J$ is a subset of $\varepsilon(i,I)$ and $i\notin J$ enforce that $J = I$, and computed and simplified the interior product in \eqref{eq:124}. As for the first summand in \eqref{eq:122}, we similarly obtain
\begin{align}
	\sum_{i}\sum_{I,J\in\mathcal{I}_{s}:i\notin I,i\in J}\Delta_{ii}\sigma(i,I)\fieldAi_J\partial_i \fieldBi_I\ebf_J\lintprod\ebf_{\varepsilon(i,I)} &= \sum_{i}\sum_{j\neq i}\sum_{K\in\mathcal{I}_{s-1}:i,j\notin K}\Delta_{ii}\sigma(i,j+K)\fieldAi_{i+K}\partial_i \fieldBi_{j+K}\ebf_{i+K}\lintprod\ebf_{i+j+K} \label{eq:125} \\
	&= \sum_{i}\sum_{j\neq i}\sum_{K\in\mathcal{I}_{s-1}:i,j\notin K}\Delta_{KK}\sigma(i,j+K)\sigma(j,i+K)\fieldAi_{i+K}\partial_i \fieldBi_{j+K}\ebf_{j} \label{eq:126} \\
	&= -\sum_{i}\sum_{j\neq i}\sum_{K\in\mathcal{I}_{s-1}:i,j\notin K}\Delta_{KK}\sigma(i,K)\sigma(j,K)\fieldAi_{i+K}\partial_i \fieldBi_{j+K}\ebf_{j}, \label{eq:127}
\end{align}
where we rewrote the summation over $I$ and $J$ in \eqref{eq:125} in terms of an index $j$ and a set $K \in \mathcal{I}_{s-1}$ such that $i,j\notin K$ and $I = \varepsilon(j,K)$ and $J = \varepsilon(i,K)$, computed the interior product in \eqref{eq:126}, and used the fact that $\sigma(i,j+K)\sigma(j,i+K) = -\sigma(i,K)\sigma(j,K)$ in \eqref{eq:127}; this latter identity follows from the fact that ordering the list $(i,j,K)$ can be done in two ways, with respective signatures $\sigma(j,K)\sigma(i,j+K)$ and $-\sigma(i,K)\sigma(j,i+K)$, whose values should coincide. Substituting $\fieldB$ for $\fieldA$ in \eqref{eq:122} and \eqref{eq:127} and grouping terms, we thus obtain
\begin{align}
	\fieldA\lintprod(\deltabf\wedge\fieldB) + \fieldB\lintprod(\deltabf\wedge\fieldA) = -\sum_{i}\sum_{j\neq i}&\sum_{K\in\mathcal{I}_{s-1}:i,j\notin K}\Delta_{KK}\sigma(i,K)\sigma(j,K)\bigl(\fieldAi_{i+K}\partial_i \fieldBi_{j+K} + \fieldBi_{i+K}\partial_i \fieldAi_{j+K}\bigr)\ebf_{j} + \notag \\
	& + \sum_{i}\sum_{I\in\mathcal{I}_{s}:i\notin I}\Delta_{ii}\Delta_{II}\partial_i (\fieldAi_I\fieldBi_I)\ebf_i. \label{eq:127b}
\end{align}

The second summation, over $i$ and $I$, in \eqref{eq:127b} can be seen to coincide with \eqref{eq:117}, apart from a factor $(-1)^s$ which cancels out with the same factor present in \eqref{eq:114}. As for the first summation, over $i$, $j$, $K$, in \eqref{eq:127b}, and again apart from common factors, each of the terms in the triple summation in \eqref{eq:127b} contains the term
\begin{equation}
	\fieldAi_{i+K}\partial_i \fieldBi_{\ell+K} + \fieldBi_{i+K}\partial_i \fieldAi_{\ell+K}, \label{eq:129}
\end{equation}
to be compared to the analogous term in the triple summation over $i$, $j$ and $K$ in \eqref{eq:119}, namely
\begin{equation}
	\partial_i\bigl(\fieldAi_{\varepsilon(\max(i,j),K)}\fieldBi_{\varepsilon(\min(i,j),K)}\bigr) = \fieldAi_{\varepsilon(\max(i,j),K)}\partial_i(\fieldBi_{\varepsilon(\min(i,j),K)}) + \fieldBi_{\varepsilon(\min(i,j),K)}\partial_i(\fieldAi_{\varepsilon(\max(i,j),K)}). \label{eq:130}
\end{equation}
Adding and subtracting some terms in \eqref{eq:130}, as well as observing that the pair of $\max(i,j)$ and $\min(i,j)$ is either $(i,j)$ or $(j,i)$, but always contains both $i$ and $j$, we obtain
\begin{align}
	\partial_i\bigl(\fieldAi_{\varepsilon(\max(i,j),K)}\fieldBi_{\varepsilon(\min(i,j),K)}\bigr) 
	&= \fieldAi_{\varepsilon(\max(i,j),K)}\partial_i(\fieldBi_{\varepsilon(\min(i,j),K)}) + \fieldBi_{\varepsilon(\min(i,j),K)}\partial_i(\fieldAi_{\varepsilon(\max(i,j),K)})  \notag \\
	&\qquad + \fieldAi_{\varepsilon(\min(i,j),K)}\partial_i(\fieldBi_{\varepsilon(\max(i,j),K)}) - \fieldAi_{\varepsilon(\min(i,j),K)}\partial_i(\fieldBi_{\varepsilon(\max(i,j),K)})  \notag \\
	&\qquad + \fieldBi_{\varepsilon(\max(i,j),K)}\partial_i(\fieldAi_{\varepsilon(\min(i,j),K)}) - \fieldBi_{\varepsilon(\max(i,j),K)}\partial_i(\fieldAi_{\varepsilon(\min(i,j),K)}) \\
	&= \fieldAi_{\varepsilon(i,K)}\partial_i(\fieldBi_{\varepsilon(j,K)}) + \fieldBi_{\varepsilon(i,K)}\partial_i(\fieldAi_{\varepsilon(j,K)})  \notag \\
	&\qquad + \fieldAi_{\varepsilon(j,K)}\partial_i(\fieldBi_{\varepsilon(i,K)}) - \fieldAi_{\varepsilon(\min(i,j),K)}\partial_i(\fieldBi_{\varepsilon(\max(i,j),K)})  \notag \\
	&\qquad + \fieldBi_{\varepsilon(j,K)}\partial_i(\fieldAi_{\varepsilon(i,K)}) - \fieldBi_{\varepsilon(\max(i,j),K)}\partial_i(\fieldAi_{\varepsilon(\min(i,j),K)}),  \label{eq:132}
\end{align}
where some of the terms in \eqref{eq:132} coincide with those in \eqref{eq:129}. We thus conclude that
\begin{align}
		(-1)^s\deltabf\lintprod(\fieldA\owedge\fieldB) &= \fieldA\lintprod(\deltabf\wedge\fieldB) + \fieldB\lintprod(\deltabf\wedge\fieldA) - R_s(\fieldA,\fieldB), \label{eq:133}
\end{align}
where the function $R_s(\fieldA,\fieldB)$ is given by		
\begin{align}
	R_s(\fieldA,\fieldB) = \sum_{i}\sum_{j\neq i}\sum_{K\in\mathcal{I}_{s-1}:i,j\notin K}&\Delta_{KK}\sigma(i,K)\sigma(j,K)\bigl( \fieldAi_{\varepsilon(j,K)}\partial_i(\fieldBi_{\varepsilon(i,K)}) - \fieldAi_{\varepsilon(\min(i,j),K)}\partial_i(\fieldBi_{\varepsilon(\max(i,j),K)}) \notag \\
	&\quad + \fieldBi_{\varepsilon(j,K)}\partial_i(\fieldAi_{\varepsilon(i,K)}) - \fieldBi_{\varepsilon(\max(i,j),K)}\partial_i(\fieldAi_{\varepsilon(\min(i,j),K)}) \bigr)\ebf_j. \label{eq:134}
\end{align}

We now get back to the remaining term in \eqref{eq:114} and again put the definition of $\odot$ in \eqref{eq:def-odot} into \eqref{eq:int-der-abset} to obtain
\begin{align}
	\deltabf\lintprod(\fieldA\odot\fieldB) 
	&= \sum_{i,j} \Delta_{ii}\Delta_{jj}\partial_i\bigl((\ebf_{\min(i,j)}\lintprod\fieldA)\cdot(\fieldB\rintprod\ebf_{\max(i,j)})\bigr)\ebf_j \\
	&= \sum_{i}\Biggl(\partial_i\bigl((\ebf_{i}\lintprod\fieldA)\cdot(\fieldB\rintprod\ebf_{i})\bigr)\ebf_i\Biggr) + \sum_i\sum_{j\neq i}\Biggl(\Delta_{ii}\Delta_{jj}\partial_i\bigl((\ebf_{\min(i,j)}\lintprod\fieldA)\cdot(\fieldB\rintprod\ebf_{\max(i,j)})\bigr)\ebf_j\Biggr), \label{eq:136}
\end{align}
where we rewrote the double summation over the pair of indices $i$ and $j$ in a convenient way. In an analogous manner to the previous analysis, using \eqref{eq:65} to evaluate the first summation over $i$ in \eqref{eq:136} yields,
\begin{align}
	\sum_{i} \partial_i\bigl((\ebf_{i}\lintprod\fieldA)\cdot(\fieldB\rintprod\ebf_{i})\bigr)\ebf_i &= (-1)^{s-1}\sum_{i} \sum_{I\in\mathcal{I}_{s}:i\in I}\Delta_{ii}\Delta_{II}\partial_i\bigl(\fieldAi_I\fieldBi_I\bigr)\ebf_i. \label{eq:137}
\end{align}
In a similar manner, we use \eqref{eq:71} to evaluate the second summation, over $i$ and $j\neq i$, in \eqref{eq:136},
\begin{align}
	\sum_{i\neq j} \Delta_{ii}\Delta_{jj}\partial_i&\bigl((\ebf_{\min(i,j)}\lintprod\fieldA)\cdot(\fieldB\rintprod\ebf_{\max(i,j)})\bigr)\ebf_j = \notag \\
	&= \sum_{i}\sum_{j\neq i}\sum_{K\in\mathcal{I}_{s-1}:i,j\notin K}\Delta_{KK}\sigma\bigl(K,\min(i,j)\bigr)\sigma\bigl(\max(i,j),K\bigr)\partial_i\bigl(\fieldAi_{\varepsilon(\min(i,j),K)}\fieldBi_{\varepsilon(\max(i,j),K)}\bigr)\ebf_j \\
	&= (-1)^{s-1}\sum_{i}\sum_{j\neq i}\sum_{K\in\mathcal{I}_{s-1}:i,j\notin K}\Delta_{KK}\sigma(K,i)\sigma(K,j)\partial_i\bigl(\fieldAi_{\varepsilon(\min(i,j),K)}\fieldBi_{\varepsilon(\max(i,j),K)}\bigr)\ebf_j, \label{eq:139} 
\end{align}
where we used that $\sigma(K,\ell) = (-1)^{\len{K}}\sigma(\ell,K)$, and that $\sigma\bigl(K,\min(i,j)\bigr)\sigma\bigl(K,\max(i,j)\bigr) = \sigma(K,i)\sigma(K,j)$ as $i\neq j$.

We proceed by relating \eqref{eq:136} to a closed-form, coordinate-free expression. First, we evaluate $\fieldA\rintprod(\deltabf\lintprod\fieldB)$, 
\begin{align}
	\fieldA\rintprod(\deltabf\lintprod\fieldB) &= \Biggl(\sum_{J\in\mathcal{I}_{s}}\fieldAi_J\ebf_J\Biggr)\rintprod\Biggl(\sum_{i,I\in\mathcal{I}_{s}:i\in I}\sigma(I\setminus i,i) \partial_i \fieldBi_I \ebf_{I\setminus i} \Biggr) \\
	&= \sum_{i,I\in\mathcal{I}_{s}:i\in I}\sum_{J\in\mathcal{I}_{s}}\sigma(I\setminus i,i)\fieldAi_J\partial_i \fieldBi_I \ebf_J\rintprod\ebf_{I\setminus i} \\
	&= \sum_{i}\sum_{I,J\in\mathcal{I}_{s}:i\in I,i\in J}\sigma(I\setminus i,i)\fieldAi_J\partial_i \fieldBi_I \ebf_J\rintprod\ebf_{I\setminus i} + \sum_{i}\sum_{I,J\in\mathcal{I}_{s}:i\in I,i\notin J}\sigma(I\setminus i,i)\fieldAi_J\partial_i \fieldBi_I \ebf_J\rintprod\ebf_{I\setminus i}, \label{eq:142}
\end{align}
where we split the summation over $J$ in two parts, depending on whether $i \in J$ or not. Evaluating the first summation over $i$ in \eqref{eq:142} yields,
\begin{align}
	\sum_{i}\sum_{I,J\in\mathcal{I}_{s}:i\in I,i\in J}\sigma(I\setminus i,i)\fieldAi_J\partial_i \fieldBi_I \ebf_J\rintprod\ebf_{I\setminus i} &= \sum_{i}\sum_{I\in\mathcal{I}_{s}:i\in I}\sigma(I\setminus i,i)\fieldAi_I\partial_i \fieldBi_I \ebf_I\rintprod\ebf_{I\setminus i} \label{eq:143} \\
	&= \sum_{i}\sum_{I\in\mathcal{I}_{s}:i\in I} \fieldAi_I\partial_i \fieldBi_I \ebf_i, \label{eq:144}
\end{align}
where we used in \eqref{eq:143} that the constraints that $I{\setminus} i$ is a subset of $J$ and $i\in J$ enforce that $J = I$, and computed and simplified the interior product in \eqref{eq:144}. As for the second summand in \eqref{eq:142}, we similarly obtain
\begin{align}
	\sum_{i}\sum_{I,J\in\mathcal{I}_{s}:i\in I,i\notin J}\sigma(I\setminus i,i)\fieldAi_J\partial_i \fieldBi_I \ebf_J\rintprod\ebf_{I\setminus i} &= \sum_{i}\sum_{\ell\neq i}\sum_{K\in\mathcal{I}_{s-1}:i\notin K}\sigma(K,i)\fieldAi_{\ell+K}\partial_i \fieldBi_{i+K} \ebf_{\ell+K}\rintprod\ebf_{K} \label{eq:145} \\
	&= \sum_{i}\sum_{\ell\neq i}\sum_{K\in\mathcal{I}_{s-1}:i\notin K}\Delta_{KK}\sigma(K,i)\sigma(K,\ell)\fieldAi_{\ell+K}\partial_i \fieldBi_{i+K} \ebf_{\ell} \label{eq:146}
\end{align}
where we rewrote the summation over $I$ and $J$ in \eqref{eq:145} in terms of an index $j$ and a set $K \in \mathcal{I}_{s-1}$ such that $i,j\notin K$ and $I = \varepsilon(i,K)$ and $J = \varepsilon(j,K)$, and computed the interior product in \eqref{eq:146}. 

Substituting $\fieldB$ for $\fieldA$ in \eqref{eq:142} and \eqref{eq:146} and grouping terms, we thus obtain
\begin{align}
	\fieldA\rintprod(\deltabf\lintprod\fieldB) + \fieldB\rintprod(\deltabf\lintprod\fieldA) = \sum_{i}\sum_{\ell\neq i}\sum_{K\in\mathcal{I}_{s-1}:i\notin K}&\Delta_{KK}\sigma(K,i)\sigma(K,\ell)\bigl(\fieldAi_{\ell+K}\partial_i \fieldBi_{i+K} + \fieldBi_{\ell+K}\partial_i \fieldAi_{i+K}\bigr) \ebf_{\ell}\notag \\ 
	&+ \sum_{i}\sum_{I\in\mathcal{I}_{s}:i\in I} \partial_i (\fieldAi_I\fieldBi_I) \ebf_i. \label{eq:147}
\end{align}
The second summation, over $i$ and $I$, in \eqref{eq:147} can be seen to coincide with \eqref{eq:137}, apart from a factor $(-1)^s$ that cancels out with the same factor present in \eqref{eq:114} and a minus sign. As for the first summation, over $i$, $j$, $K$, in \eqref{eq:147}, and again apart from common factors, each of the terms in the triple summation in \eqref{eq:147} contains the term
\begin{equation}
	\fieldAi_{\ell+K}\partial_i \fieldBi_{i+K} + \fieldBi_{\ell+K}\partial_i \fieldAi_{i+K}, \label{eq:149}
\end{equation}
to be compared to the analogous term in the triple summation over $i$, $j$ and $K$ in \eqref{eq:139}, namely,
\begin{equation}
	\partial_i\bigl(\fieldAi_{\varepsilon(\min(i,j),K)}\fieldBi_{\varepsilon(\max(i,j),K)}\bigr) = \fieldAi_{\varepsilon(\min(i,j),K)}\partial_i(\fieldBi_{\varepsilon(\max(i,j),K)}) + \fieldBi_{\varepsilon(\max(i,j),K)}\partial_i(\fieldAi_{\varepsilon(\min(i,j),K)}). \label{eq:150}
\end{equation}
Adding and subtracting some terms in \eqref{eq:150}, as well as observing that the pair of $\max(i,j)$ and $\min(i,j)$ is either $(i,j)$ or $(j,i)$, but always contains both $i$ and $j$, we obtain
\begin{align}
	\partial_i\bigl(\fieldAi_{\varepsilon(\min(i,j),K)}\fieldBi_{\varepsilon(\max(i,j),K)}\bigr) &= \fieldAi_{\varepsilon(\min(i,j),K)}\partial_i(\fieldBi_{\varepsilon(\max(i,j),K)}) + \fieldBi_{\varepsilon(\max(i,j),K)}\partial_i(\fieldAi_{\varepsilon(\min(i,j),K)})  \notag \\
	&\qquad + \fieldAi_{\varepsilon(\max(i,j),K)}\partial_i(\fieldBi_{\varepsilon(\min(i,j),K)}) - \fieldAi_{\varepsilon(\max(i,j),K)}\partial_i(\fieldBi_{\varepsilon(\min(i,j),K)})  \notag \\
	&\qquad + \fieldBi_{\varepsilon(\min(i,j),K)}\partial_i(\fieldAi_{\varepsilon(\max(i,j),K)}) - \fieldBi_{\varepsilon(\min(i,j),K)}\partial_i(\fieldAi_{\varepsilon(\max(i,j),K)}) \\
	&= \fieldAi_{\varepsilon(i,K)}\partial_i(\fieldBi_{\varepsilon(j,K)}) + \fieldBi_{\varepsilon(i,K)}\partial_i(\fieldAi_{\varepsilon(j,K)})  \notag \\
	&\qquad + \fieldAi_{\varepsilon(j,K)}\partial_i(\fieldBi_{\varepsilon(i,K)}) - \fieldAi_{\varepsilon(\max(i,j),K)}\partial_i(\fieldBi_{\varepsilon(\min(i,j),K)})  \notag \\
	&\qquad + \fieldBi_{\varepsilon(j,K)}\partial_i(\fieldAi_{\varepsilon(i,K)}) - \fieldBi_{\varepsilon(\min(i,j),K)}\partial_i(\fieldAi_{\varepsilon(\max(i,j),K)}), \label{eq:152}
\end{align}
where some of the terms in \eqref{eq:152} coincide with those in \eqref{eq:149}. We may therefore conclude that
\begin{align}
		(-1)^s\deltabf\lintprod(\fieldA\odot\fieldB) &= -\fieldA\rintprod(\deltabf\lintprod\fieldB) - \fieldB\rintprod(\deltabf\lintprod\fieldA) - Q_s(\fieldA,\fieldB), \label{eq:153}
\end{align}
where the function $Q_s(\fieldA,\fieldB)$ is given by		
\begin{align}
	Q_s(\fieldA,\fieldB) = \sum_{i}\sum_{j\neq i}\sum_{K\in\mathcal{I}_{s-1}:i,j\notin K} &\Delta_{KK}\sigma(i,K)\sigma(j,K)\bigl( \fieldAi_{\varepsilon(i,K)}\partial_i(\fieldBi_{\varepsilon(j,K)}) - \fieldAi_{\varepsilon(\max(i,j),K)}\partial_i(\fieldBi_{\varepsilon(\min(i,j),K)}) \notag \\
	&\quad  + \fieldBi_{\varepsilon(i,K)}\partial_i(\fieldAi_{\varepsilon(j,K)}) - \fieldBi_{\varepsilon(\min(i,j),K)}\partial_i(\fieldAi_{\varepsilon(\max(i,j),K)}) 
\bigr)\ebf_j. \label{eq:154}
\end{align}
It remains to verify that $R_s(\fieldA,\fieldB) + Q_s(\fieldA,\fieldB) = 0$, where $R_s(\fieldA,\fieldB)$ was given in \eqref{eq:134}. This condition is satisfied if
\begin{align}
&\fieldAi_{\varepsilon(j,K)}\partial_i(\fieldBi_{\varepsilon(i,K)}) - \fieldAi_{\varepsilon(\min(i,j),K)}\partial_i(\fieldBi_{\varepsilon(\max(i,j),K)}) 
	+ \fieldBi_{\varepsilon(j,K)}\partial_i(\fieldAi_{\varepsilon(i,K)}) - \fieldBi_{\varepsilon(\max(i,j),K)}\partial_i(\fieldAi_{\varepsilon(\min(i,j),K)}) \notag \\
	 &+\fieldAi_{\varepsilon(i,K)}\partial_i(\fieldBi_{\varepsilon(j,K)}) - \fieldAi_{\varepsilon(\max(i,j),K)}\partial_i(\fieldBi_{\varepsilon(\min(i,j),K)}) + \fieldBi_{\varepsilon(i,K)}\partial_i(\fieldAi_{\varepsilon(j,K)}) - \fieldBi_{\varepsilon(\min(i,j),K)}\partial_i(\fieldAi_{\varepsilon(\max(i,j),K)}) = 0,
\end{align}
as can be verified for both possible orderings of the pair $(i,j)$, that is $\max(i,j) = i$, $\max(i,j) = j$. Therefore, we may combine \eqref{eq:133} and \eqref{eq:153} into \eqref{eq:114} into the final, closed-form, coordinate-free expression in \eqref{eq:int-der-abset-cf}, that is
\begin{equation}
	\deltabf\lintprod\abset = \fieldA\lintprod(\deltabf\wedge\fieldB) + \fieldB\lintprod(\deltabf\wedge\fieldA) -\fieldA\rintprod(\deltabf\lintprod\fieldB)  - \fieldB\rintprod (\deltabf\lintprod\fieldA).
\end{equation}

\section{Examples and applications}
\label{sec:applications}
\subsection{Conformal invariance}

Conformal invariance of a field theory is related to the vanishing of the trace of the stress-energy momentum tensor \cite[Sect.\ 4.2]{diFrancesco1997conformalFieldTheory}. From the definition of the trace, we may compute the trace of $\abset$ from \eqref{eq:abset_ii}:
\begin{align}
	\Tr \abset &= \sum_{i}\Delta_{ii}\abseti_{ii} \label{eq:trace-abset-0} \\
	&= \sum_{i}\Biggl(\sum_{K\in{\mathcal{I}_{s}}:i\notin K}\Delta_{KK}\fieldAi_{K}\fieldBi_{K}-\sum_{K\in{\mathcal{I}_{s}}:i\in K}\Delta_{KK}\fieldAi_{K}\fieldBi_{K}\Biggr) \label{eq:trace-abset-1} \\
	&= \sum_{K\in{\mathcal{I}_{s}}}\Biggl(\sum_{i:i\notin K}\Delta_{KK}\fieldAi_{K}\fieldBi_{K}-\sum_{i:i\in K}\Delta_{KK}\fieldAi_{K}\fieldBi_{K}\Biggr) \label{eq:trace-abset-2} \\
	&= \sum_{K\in{\mathcal{I}_{s}}}\bigl((k+n-s)\Delta_{KK}\fieldAi_{K}\fieldBi_{K}-s\Delta_{KK}\fieldAi_{K}\fieldBi_{K}\bigr) \label{eq:trace-abset-3} \\
	&= (k+n-2s)(\fieldA\cdot\fieldB), \label{eq:trace-abset-4}
\end{align}
where \eqref{eq:trace-abset-0} follows from the definition of trace, in \eqref{eq:trace-abset-1} we replaced the coefficient $\abseti_{ii}$ by its formula in \eqref{eq:abset_ii}, we reversed the summation order in \eqref{eq:trace-abset-2}, we counted the number of indices $i$ appearing in each summation over $i$ for fixed $s$-tuple $K$ in \eqref{eq:trace-abset-3}, and we finally used the definition of $\fieldA\cdot\fieldB$ in \eqref{eq:trace-abset-4}. From \eqref{eq:trace-abset-4}, we obtain
\begin{equation}\label{eq:trace-totalset}
	\Tr\semt{sys} = \sum_{\fieldA,\fieldB}\gamma_{\fieldA,\fieldB}\bigl(k+n-2\gr(\fieldA)\bigr)(\fieldA\cdot\fieldB),
\end{equation}
for the action given in \eqref{eq:sem_total}.
The formula in \eqref{eq:trace-totalset} is a function of the fields appearing explicitly in the Lagrangian density only. In particular, if $\fieldA$ or $\fieldB$ have some internal structure, e.\ g.,\ $\fieldA = \deltabf\wedge\vp$ in generalized electromagnetism or Yang-Mills fields, this vector potential may be bypassed. The same principle holds for the following examples, and need not explicitly consider the internal structure of the Lagrangian density terms in our analysis.

\subsection{Scalar field}
\newcommand{\scf}{\phi}

In flat $(k,n)$-dimensional space-time, the Lagrangian density $\ld_\text{free-scalar}$ of a free scalar field $\scf$ is given in \eqref{eq:lagrangian_scalar_free}, 
so we can make the identification $\fieldA = \fieldB = \deltabf\wedge\scf = \sum_i\Delta_{ii}\partial_i\scf\,\ebf_i$, and $\gr(\fieldA) = \gr(\fieldB) = 1$. Taking into account the multiplicative factor $\frac{1}{2}$ in $\ld$, the on-diagonal component of the tensor \eqref{eq:abset-uij} in \eqref{eq:abset_ii} can be directly evaluated as:
\begin{align}
	\semti{free-scalar}_{ii} &= \frac{1}{2}\Delta_{ii}\Biggl(-\Delta_{ii}\fieldAi_{i}^2 + \sum_{j:i\neq j}\Delta_{jj}\fieldAi_{j}^2\Biggr) \\
	&= -\frac{1}{2}(\partial_i \scf)^2 + \frac{1}{2}\sum_{j:i\neq j}\Delta_{ii}\Delta_{jj}(\partial_j \scf)^2.
\end{align}
As for the off-diagonal terms, using \eqref{eq:abset_ij} and taking into account that $\mathcal{I}_{0}$ contains only the empty set $\O$, we obtain
\begin{align}
	\semti{free-scalar}_{ij} &= -\fieldAi_{i}\fieldAi_{j} \\
	&= -\Delta_{ii}\Delta_{jj}(\partial_i \scf)(\partial_j \scf).
\end{align}

The interior derivative of the tensor $\semt{scalar}$ is computed from \eqref{eq:int-der-abset-cf}, that is
\begin{align}
	\deltabf\lintprod\semt{free-scalar} &= \fieldA\lintprod(\deltabf\wedge\fieldA) - \fieldA\rintprod(\deltabf\lintprod\fieldA) \label{eq:164} \\
	&= - (\deltabf\wedge\scf)\bigl(\deltabf\lintprod(\deltabf\wedge\scf)\bigr) \label{eq:165} \\
	&= - \bigl((\deltabf\cdot\deltabf)\scf\bigr)(\deltabf\wedge\scf), \label{eq:166}
\end{align}
where we took into account the multiplicative factor $\frac{1}{2}$ in the Lagrangian density and used that $\fieldA = \fieldB$ in \eqref{eq:164},  used that $\deltabf\wedge(\deltabf\wedge\phi) = 0$ (cf.\ \cite[Eq.\ (32)]{colombaro2020generalizedMaxwellEquations}) and that $\deltabf\lintprod\fieldA$ is a scalar in \eqref{eq:165}, and the identity \cite[Eq.\ (34)]{colombaro2020generalizedMaxwellEquations} in \eqref{eq:166} to rewrite the term $\deltabf\lintprod(\deltabf\wedge\scf)$ as  $(\deltabf\cdot\deltabf)\scf$.

\subsection{Electromagnetism}

In flat $(k,n)$-dimensional space-time, the Lagrangian density $\ld_\text{free-gem}$ of a free generalized electromagnetic field $\mf$, a multivector field of grade $r$, with vector potential $\vp$, such that $\mf = \deltabf\wedge\vp$, is given in \eqref{eq:lagrangian_gem_free}.
From \eqref{eq:abset-oo}, we may express the stress-energy-momentum tensor $\semt{free-gem}$ for any values of $r$, $k$ and $n$ as
\footnote{We have moved a factor $\frac{1}{2}$ from the definition of $\mf\odot\mf$ and $\mf\owedge\mf$ in \cite{colombaro2020generalizedMaxwellEquations}. Also, the proof in Appendix A.2 of \cite{colombaro2020generalizedMaxwellEquations} should be amended as of \eqref{eq:133} and \eqref{eq:153} in this paper; the final formula for the interior derivative given in Appendix A.2 of \cite{colombaro2020generalizedMaxwellEquations} remains unchanged.}
\begin{equation}
\semt{free-gem} = -\frac{1}{2}(\mf\owedge\mf+\mf\odot\mf),
\end{equation}
with on- and off-diagonal components respective given by \eqref{eq:abset_ii_intro} and \eqref{eq:abset_ij_intro},
\begin{align}
	\semti{free-gem}_{ii} &= \frac{(-1)^{r}}{2}\Delta_{ii}\Biggl(\sum_{I\in\mathcal{I}_{r}: i \in I}\mfi_I^2\Delta_{II} - \sum_{I\in\mathcal{I}_{r}: i \notin I}\mfi_{I}^2\Delta_{II}\Biggr) \\
	\semti{free-gem}_{ij} &= -\sum_{L\in\mathcal{I}_{r-1}:i,j\neq L}\sigma(L,i)\sigma(j,L)\mfi_{i+L} \mfi_{j+L} \Delta_{LL},
\end{align}
in alignment with \cite{colombaro2020generalizedMaxwellEquations} and with the stress-energy tensor for standard electromagnetism with bivectors, or the Faraday tensor ($r = 2$, $k = 1$, $n = 3$) \cite[Sect.\ 33]{landau1982classicalTheoryFields}, \cite[Sect.\ 12.10]{jackson1999classicalElectrodynamics}.

While the physical interpretation remains open, we may apply \eqref{eq:abset-oo} to find the stress-energy-momentum tensor of the interaction Lagrangian density  $\sd\cdot\vp$, which denote as $\semt{int-gem}$, as 
\begin{equation}
	\semt{int-gem} = (-1)^{r-1}(\sd\owedge\vp+\sd\odot\vp),
\end{equation}
with on- and off-diagonal components respective given by \eqref{eq:abset_ii_intro} and \eqref{eq:abset_ij_intro},
\begin{align}
	\semti{int-gem}_{ii} &= \Delta_{ii}\Biggl(\sum_{K\in{\mathcal{I}_{s}}:i\notin K}\Delta_{KK}\sdi_{K}\vpi_{K}-\sum_{K\in{\mathcal{I}_{s}}:i\in K}\Delta_{KK}\sdi_{K}\vpi_{K}\Biggr),  \\
	\semti{int-gem}_{ij} &= -\sum_{K\in{\mathcal{I}_{s-1}}:i,j\notin K}\Delta_{KK}\sigma\bigl(\varepsilon(i,K)_{i\leftrightarrow j}\bigr)\bigl(\sdi_{\varepsilon(i,K)}\vpi_{\varepsilon(j,K)}+\vpi_{\varepsilon(i,K)}\sdi_{\varepsilon(j,K)}\bigr). 
\end{align}

The transfer of energy-momentum is described by the interior derivative of the stress-energy-momentum tensor. Applying \eqref{eq:int-der-abset-cf} to $\semt{free-gem}$ and $\semt{int-gem}$, respectively, gives:
\begin{gather}
	\deltabf\lintprod\semt{free-gem} = (-1)^{r-1} \bigl(\mf\lintprod(\deltabf\wedge\mf) - \mf\rintprod(\deltabf\lintprod\mf)\bigr) \label{eq:188} \\
	\deltabf\lintprod\semt{int-gem} = \sd\lintprod(\deltabf\wedge\vp) + \vp\lintprod(\deltabf\wedge\sd) -\sd\rintprod(\deltabf\lintprod\vp)  - \vp\rintprod (\deltabf\lintprod\sd). \label{eq:189}
\end{gather}
Setting now $\mf = \deltabf\wedge\vp$, $\deltabf\lintprod\mf = \sd$ and $\deltabf\wedge\mf = 0$ in \eqref{eq:188}--\eqref{eq:189}, and combining the resulting expressions yields
\begin{align}
	\deltabf\lintprod(\semt{free-gem} + \semt{int-gem}) &= (-1)^{r} \mf\rintprod(\deltabf\lintprod\mf) + \sd\lintprod\mf + \vp\lintprod(\deltabf\wedge\sd) -\sd\rintprod(\deltabf\lintprod\vp)  - \vp\rintprod (\deltabf\lintprod\sd) \\
	&= (-1)^{r}\mf\rintprod\sd + \sd\lintprod\mf + \vp\lintprod(\deltabf\wedge\sd) -\sd\rintprod(\deltabf\lintprod\vp) \label{eq:191} \\
	&= -\sd\lintprod\mf + \sd\lintprod\mf + \vp\lintprod(\deltabf\wedge\sd) -\sd\rintprod(\deltabf\lintprod\vp) \label{eq:192} \\
	&= \vp\lintprod(\deltabf\wedge\sd) -\sd\rintprod(\deltabf\lintprod\vp), \label{eq:193}
\end{align}
where we used in \eqref{eq:191} the Maxwell equation $\deltabf\lintprod\mf = \sd$ and the continuity equation for the current $\deltabf\lintprod\sd = 0$, and the identity $\mf\rintprod\sd = (-1)^{r-1}\sd\lintprod\mf$ in \eqref{eq:192}. The generalized Lorentz force density $\fb = \Jb\lintprod \Fb$ cancels out from the interior derivative as $	\fb + \deltabf\lintprod \semt{free-gem} = 0$. However, the physical interpretation of the terms  in \eqref{eq:193}, including their Gauge invariance and possible connection to the tensor of the matter fields, remains open.

\subsection{Yang-Mills fields}

In flat $(1,3)$-dimensional space-time, the Lagrangian density $\ld_\text{free-ym}$ of a free Yang-Mills field $\mf$, a Lie-algebra valued bivector field, with connection $\vp$, is given in \eqref{eq:lagrangian_ym_free_1}--\eqref{eq:lagrangian_ym_free_2}.
From \eqref{eq:abset-oo}, we may directly express the stress-energy-momentum tensor $\semt{free-ym}$ as
\begin{align}
	\semt{free-ym} &= -\frac{1}{2g^2}\Tr(\mf\owedge\mf+\mf\odot\mf) \\
	&= -\frac{1}{2g^2}\sum_a (\mf^a\owedge\mf^a+\mf^a\odot\mf^a),
\end{align}
with on- and off-diagonal components respective given by \eqref{eq:abset_ii_intro} and \eqref{eq:abset_ij_intro}, 
\begin{gather}
	\semti{free-ym}_{ii} = \frac{(-1)^{r}}{4g^2}\Delta_{ii}\sum_a\Biggl(\sum_{I\in\mathcal{I}_{r}: i \in I}(\mfi_I^a)^2\Delta_{II} - \sum_{I\in\mathcal{I}_{r}: i \notin I}(\mfi_{I}^a)^2\Delta_{II}\Biggr) \\
	\semti{free-ym}_{ij} = -\frac{1}{2g^2}\sum_a\sum_{L\in\mathcal{I}_{r-1}:i,j\neq L}\sigma(L,i)\sigma(j,L)\mfi_{i+L}^a \mfi_{j+L}^a \Delta_{LL},
\end{gather}
in agreement with \cite[Eq.\ (2.11)]{Blaschke_2016} and \cite[Eq.\ (46)]{montesinos2006symmetricTensor}.
The transfer of energy-momentum from the Yang-Mills field is described by the interior derivative of $\semt{free-ym}$ given in \eqref{eq:int-der-abset-cf}.
 
\subsection{Proca field}

The Lagrangian density $\ld_\text{proca}$ of a Proca field $\vp$ is given in \eqref{eq:lagrangian_proca}, with $\mf = \deltabf\wedge\vp$. We obtain its stress-energy-momentum tensor by adding to the tensor $\semt{free-gem}$ with $r = 2$ another tensor $\semt{mass-proca}$ for the mass terms, 
\begin{equation}
	\semt{mass-proca} = -\frac{1}{2}m^2(\vp\owedge\vp+\vp\odot\vp),
\end{equation}
with on- and off-diagonal components respectively given by \eqref{eq:abset_ii_intro} and \eqref{eq:abset_ij_intro},
\begin{gather}
	\semti{mass-proca}_{ii} = \Delta_{ii}\frac{m^2}{2}\Biggl(\sum_{K\in{\mathcal{I}_{s}}:i\notin K}\Delta_{KK}\vpi_{K}^2-\sum_{K\in{\mathcal{I}_{s}}:i\in K}\Delta_{KK}\vpi_{K}^2\Biggr),  \\
	\semti{mass-proca}_{ij} = -m^2\sum_{K\in{\mathcal{I}_{s-1}}:i,j\notin K}\Delta_{KK}\sigma\bigl(\varepsilon(i,K)_{i\leftrightarrow j}\bigr)\vpi_{\varepsilon(i,K)}\vpi_{\varepsilon(j,K)},
\end{gather} 
in agreement with \cite[Eq.\ (62)]{montesinos2006symmetricTensor}. Applying \eqref{eq:int-der-abset-cf} to determine the interior derivative of the stress-energy-momentum tensor to the sum $\semt{free-proca} = \semt{free-gem} + \semt{mass-proca}$, together with \eqref{eq:188} after setting $\deltabf\wedge\mf = 0$, gives 
\begin{align}
	\deltabf\lintprod\semt{free-proca} &= \mf\rintprod(\deltabf\lintprod\mf) + m^2\vp\lintprod(\deltabf\wedge\vp)  -m^2\vp\rintprod(\deltabf\lintprod\vp) \\
	&= \mf\rintprod(\deltabf\lintprod\mf) + m^2\vp\lintprod\mf  -m^2\vp\rintprod(\deltabf\lintprod\vp) \label{eq:201} \\
	&= m^2\mf\rintprod\vp + m^2\vp\lintprod\mf  - m^2\vp\rintprod(\deltabf\lintprod\vp) \label{eq:202} \\
	&= -m^2\vp\rintprod(\deltabf\lintprod\vp), \label{eq:203}
\end{align}
where we used in \eqref{eq:201} the definition $\mf = \deltabf\wedge\vp$, in \eqref{eq:202} the equation $\deltabf\lintprod\mf = m^2\vp$, and the identity $\mf\rintprod\vp = -\vp\lintprod\mf$, obtained from the Euler-Lagrange equations of the system, in \eqref{eq:203}, as $\mf$ is a bivector. While the interior derivative of the free (generalized) electromagnetic field tensor \eqref{eq:188} vanishes in the absence of interaction with a current density $\sd$, the interior derivative of the Proca field tensor does not vanish unless the Lorenz gauge condition, $\deltabf\lintprod\vp$, holds, thereby breaking the gauge invariance.

\subsection{Conclusions and future work}

In this paper, we have provided an exterior-algebraic derivation of the symmetric stress-energy-momentum tensor that naturally appears in field theories from the invariance of the action of a closed physical system to infinitesimal space-time translations. Our focus lies on Lagrangian densities that are expressed as the dot product of two multivector fields, e.\ g.,\  scalar or gauge fields, in flat space-time. The analysis covers a number of interesting cases, such as electromagnetic fields, Proca fields or Yang-Mills fields, and it leaves out the relevant case of spinor matter fields. An extension to spinor fields is left for future work and will be reported elsewhere. 
Our formalism allows us to calculate the tensor and its interior derivative, not only for free fields but also for the interaction terms appearing in the action. It would be interesting to relate the interior derivative of the stress-energy-tensor associated with spinor fields, e.g.,\ that of the electron, to the interior derivative of the tensor associated with the interaction Lagrangian density.

Finally, while we have considered invariance of the action to  space-time translations, it would be worthwhile considering the full Poincar\'e group, i.e.,\ including Lorentz transformations. A study of this case would lead to an exterior-algebraic characterization of the angular momentum tensor, extending the analysis presented in this paper. Moreover, considering the conserved charges associated to the stress-energy-momentum or angular momentum tensors would lead to a redefinition of energy-momentum and angular momentum in general flat space-times.

\appendix

\chapter{Proof of $\sigma\bigl(\varepsilon(i,K)_{i\leftrightarrow j}\bigr) =  \sigma\bigl(\varepsilon(j,K)_{j\leftrightarrow i}\bigr) $ in~\eqref{eq:switch-ij} and related expressions}
\label{sec:symmetry_ij}

With no real loss of generality, assume that $i < j$ and let $K$ be an index set that does not include $i$ and $j$. We can then write the set $K$ as the union of three disjoint subsets as depicted in~\figurename~\ref{fig:proof-182}. From the graphic representation in \ref{fig:proof-182}, we can express some signatures as follows:
\begin{gather}
	\sigma(K,i) = \sigma(K_3,i)\sigma(K_2,i), \label{eq:app-205}	\\
	\sigma(j,K) = \sigma(j,K_1)\sigma(j,K_2), \label{eq:app-206} \\
	\sigma(i, j+K) = \sigma(i,K_1),	 \label{eq:app-207} \\
	\sigma(i+K,j) = \sigma(K_3,j), \label{eq:app-208} \\
	\sigma\bigl(\varepsilon(i,K)_{i\leftrightarrow j}\bigr) = \sigma(j,K_2). \label{eq:app-209}
\end{gather}
As the permutations on the right-hand sides of \eqref{eq:app-205}--\eqref{eq:app-209} represent the indices $i$ and $j$ going through fixed subsets,  we may evaluate the following signatures:
\begin{gather}
	\sigma(i,K_1) = \sigma(j,K_1) = (-1)^{\len{K_1}}, \label{eq:app-210} \\
	\sigma(j,K_2) = \sigma(K_2,i) = (-1)^{\len{K_2}}, \label{eq:app-211} \\
	\sigma(K_3,j) = \sigma(K_3,i) = (-1)^{\len{K_3}}. \label{eq:app-212}
\end{gather}
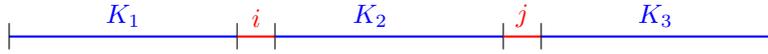
\begin{figure}[!htb]
\centering
\begin{tikzpicture}
\draw [blue, thick] (-3,4) node[black] {$\vert$} -- (0,4) node[black] {$\vert$};
\node[blue, above] at (-1.5,4) {$K_1$};
\draw [red, thick] (0,4) -- (0.5,4) node[black] {$\vert$};
\node[red, above] at (0.25,4) {$i$};
\draw [blue, thick] (0.5,4)  -- (3.5,4) node[black] {$\vert$};
\node[blue, above] at (1.75,4) {$K_2$};
\draw [red, thick] (3.5,4) -- (4,4) node[black] {$\vert$};
\node[red, above] at (3.75,4) {$j$};
\draw [blue, thick] (4,4)  -- (7,4) node[black] {$\vert$};
\node[blue, above] at (5.5,4) {$K_3$};
\end{tikzpicture}
\caption{Characterization of the set $K$ as union of three subsets, for $i<j$.}
\label{fig:proof-182}
\end{figure}

As the indices $i$ and $j$ are separated by the set $K_2$, 
the permutation that orders the set $\varepsilon(i,K)$ when $i$ is replaced by $j$ simply has to rearrange the set $(j,K_2)$, 
\begin{equation}
	\sigma\bigl(\varepsilon(i,K)_{i\leftrightarrow j}\bigr) = \sigma(j,K_2) = (-1)^{\vert K_2\vert}. \label{eq:proof-line-ij}
\end{equation}
Similarly, the permutation that orders the set $\varepsilon(j,K)$ when $j$ is replaced by $i$ has to rearrange the set $(K_2,i)$, 
\begin{equation}
	\sigma\bigl(\varepsilon(j,K)_{j\leftrightarrow i}\bigr) = \sigma(K_2, i) = (-1)^{\vert K_2\vert}. \label{eq:proof-line-ji}
\end{equation} 
As Eqs.~\eqref{eq:proof-line-ij} and~\eqref{eq:proof-line-ji} coincide, and the reasoning is unaffected if $i > j$ and Eq.~\eqref{eq:switch-ij} is proved, 
\begin{equation}
	\sigma\bigl(\varepsilon(i,K)_{i\leftrightarrow j}\bigr) =  \sigma\bigl(\varepsilon(j,K)_{j\leftrightarrow i}\bigr). \label{eq:app-215}
\end{equation}	

In fact, one can prove two alternative characterizations of $\sigma\bigl(\varepsilon(j,K)_{j\leftrightarrow i}\bigr)$, as we do next. Assume again with no loss of generality that $i < j$. First, we compute 
\begin{align}
	(-1)^{\len{K}}\sigma\bigl(\varepsilon(i,K)_{i\leftrightarrow j}\bigr) & = (-1)^{\len{K}}(-1)^{\len{K_2}}\\
	&= (-1)^{\len{K_1}}(-1)^{\len{K_3}}. \label{eq:app-217}
\end{align}
Then, we can verify by using \eqref{eq:app-205}--\eqref{eq:app-209}, \eqref{eq:app-210}--\eqref{eq:app-212}, and \eqref{eq:app-217} that
\begin{align}
	\sigma(K,i)\sigma(j,K) &= \sigma(K_3,i)\sigma(K_2,i)\sigma(j,K_1)\sigma(j,K_2) \\
	&= (-1)^{\len{K_3}}(-1)^{\len{K_2}}(-1)^{\len{K_1}}(-1)^{\len{K_2}} \\
	&= (-1)^{\len{K_3}}(-1)^{\len{K_1}} \\
	&= (-1)^{\len{K}}\sigma\bigl(\varepsilon(i,K)_{i\leftrightarrow j}\bigr). \label{eq:app-221}
\end{align}
And similarly, by using \eqref{eq:app-205}--\eqref{eq:app-209}, \eqref{eq:app-210}--\eqref{eq:app-212} and \eqref{eq:app-217} we find that
\begin{align}
	\sigma(i,j+K)\sigma(i+K,j) &= \sigma(i,K_1)\sigma(K_3,j) \\
	&= (-1)^{\len{K_1}}(-1)^{\len{K_3}} \\
	&= (-1)^{\len{K}}\sigma\bigl(\varepsilon(i,K)_{i\leftrightarrow j}\bigr). \label{eq:app-224}
\end{align}
If we carry out the analysis for $i>j$, we obtain the same expressions in \eqref{eq:app-221} and  \eqref{eq:app-224}.

%
%
%
%
%
%

\chapter{Proof of Leibniz rule for mixed product}\label{sec:leibniz-mixed}
	
We prove~\eqref{eq:67} starting by the evaluation of the left-hand side
\begin{align}
	\deltabf\cdot(\xbpert\lintprod\Tb) &= 
	\biggl( \sum_i \Delta_{ii}\partial_i \ebf_i
	\biggr) \cdot
	\Biggl( \biggl( \sum_j \xpert[j] \ebf_j\Bigr) \lintprod \Bigl( \sum_{\ell \le m} T_{\ell m} \ubf_{\ell m} \biggr)
	\Biggr)
	\\
	&= 	\biggl( \sum_i \Delta_{ii}\partial_i \ebf_i
	\biggr) \cdot
	\biggl( \sum_{j,\ell} \xpert[j] T_{\varepsilon(\ell,j)}\Delta_{jj} \ebf_\ell\biggr) \\
	&= \sum_{i,j} \Delta_{jj} \partial_i\bigl(\xpert[j] T_{\varepsilon(i,j)}\bigr). \label{app:eq-227}
\end{align}

Then, on the right-hand side of~\eqref{eq:67}, the first term can be similarly expressed as
\begin{align}
\xbpert \cdot (\deltabf \lintprod \Tb) &=
\biggl( \sum_{j} \xpert[j] \ebf_j \biggr) \cdot \Biggl( \biggl(  \sum_i \Delta_{ii}\partial_i \ebf_i \Bigr) \lintprod \Bigl( \sum_{\ell \le m} T_{\ell m} \ubf_{\ell m} \biggr) \Biggr) \\
	&= \Bigl( \sum_{j} \xpert[j] \ebf_j \Bigr) \cdot \Bigl( \sum_{i, \ell} \partial_i T_{i\ell} \ebf_\ell \Bigr) =
\sum_{i,j} \Delta_{jj} \xpert[j] \partial_iT_{\varepsilon(i,j)}. \label{app:eq-232}
\end{align}
Finally, the second term on the right-hand side of~\eqref{eq:67} can be expanded using~\eqref{eq:dot_tensor_symm} as
\begin{align}
	(\deltabf\otimes\xbpert)\cdot\Tb &= \biggl( \sum_{i,j} \Delta_{ii} \partial_i\xpert[j] \ebf_i \otimes \ebf_j  \biggr) \cdot \biggl( \sum_{\ell \le m}T_{\ell m} \ubf_{\ell m} \biggr) \\
	&= \sum_{i,j} \Delta_{jj} T_{\min(i,j),\max(i,j)} \partial_i\xpert[j] \\
	&= \sum_{i,j} \Delta_{jj} T_{\varepsilon(i,j)} \partial_i\xpert[j]. \label{app:eq-235}
\end{align}
Combining the three expressions in \eqref{app:eq-227}, \eqref{app:eq-232} and \eqref{app:eq-235} proves the generalized Leibniz rule in~\eqref{eq:67}.

\bibliographystyle{IEEEtran}	
\bibliography{physics.bib}

\begin{thebibliography}{10}
\providecommand{\url}[1]{#1}
\csname url@samestyle\endcsname
\providecommand{\newblock}{\relax}
\providecommand{\bibinfo}[2]{#2}
\providecommand{\BIBentrySTDinterwordspacing}{\spaceskip=0pt\relax}
\providecommand{\BIBentryALTinterwordstretchfactor}{4}
\providecommand{\BIBentryALTinterwordspacing}{\spaceskip=\fontdimen2\font plus
\BIBentryALTinterwordstretchfactor\fontdimen3\font minus
  \fontdimen4\font\relax}
\providecommand{\BIBforeignlanguage}[2]{{%
\expandafter\ifx\csname l@#1\endcsname\relax
\typeout{** WARNING: IEEEtran.bst: No hyphenation pattern has been}%
\typeout{** loaded for the language `#1'. Using the pattern for}%
\typeout{** the default language instead.}%
\else
\language=\csname l@#1\endcsname
\fi
#2}}
\providecommand{\BIBdecl}{\relax}
\BIBdecl

\bibitem{landau1982classicalTheoryFields}
L.~D. Landau and E.~M. Lifshitz, \emph{The Classical Theory of Fields},
  4th~ed., ser. Course of Theoretical Physics.\hskip 1em plus 0.5em minus
  0.4em\relax Butterworth-Heinemann, 1987, vol.~2.

\bibitem{jackson1999classicalElectrodynamics}
J.~D. Jackson, \emph{Classical Electrodynamics}, 3rd~ed.\hskip 1em plus 0.5em
  minus 0.4em\relax John Wiley \& Sons, 1999.

\bibitem{Forger_2004}
\BIBentryALTinterwordspacing
M.~Forger and H.~R\"{o}mer, ``Currents and the energy-momentum tensor in
  classical field theory: a fresh look at an old problem,'' \emph{Annals of
  Physics}, vol. 309, no.~2, pp. 306--389, Feb 2004. [Online]. Available:
  \url{http://dx.doi.org/10.1016/j.aop.2003.08.011}
\BIBentrySTDinterwordspacing

\bibitem{Blaschke_2016}
\BIBentryALTinterwordspacing
D.~N. Blaschke, F.~Gieres, M.~Reboud, and M.~Schweda, ``The energy-momentum
  tensor(s) in classical gauge theories,'' \emph{Nuclear Physics B}, vol. 912,
  pp. 192--223, Nov 2016. [Online]. Available:
  \url{http://dx.doi.org/10.1016/j.nuclphysb.2016.07.001}
\BIBentrySTDinterwordspacing

\bibitem{maggiore2005modernIntroductionQFT}
M.~Maggiore, \emph{{A modern introduction to quantum field theory}}, ser.
  Oxford Master Series in Statistical, Computational, and Theoretical
  Physics.\hskip 1em plus 0.5em minus 0.4em\relax Oxford: Oxford Univ. Press,
  2005.

\bibitem{diFrancesco1997conformalFieldTheory}
P.~Di~Francesco, P.~Mathieu, and D.~S\'en\'echal, \emph{{Conformal Field
  Theory}}, ser. Graduate texts in contemporary physics.\hskip 1em plus 0.5em
  minus 0.4em\relax New York, NY: Springer, 1997.

\bibitem{gotay1992stressEnergyMomentumTensors}
M.~J. Gotay and J.~E. Marsden, ``Stress-energy-momentum tensors and the
  belinfante-rosenfeld formula,'' \emph{Contemp. Math.}, vol. 132, pp.
  367--392, 1992.

\bibitem{voicu2015energy-momentumTensors}
N.~Voicu, ``Energy-momentum tensors in classical field theories -- {A} modern
  perspective,'' \emph{International Journal of Geometric Methods in Modern
  Physics}, vol.~13, p. 1640001, 2015.

\bibitem{colombaro2019introductionSpaceTimeExteriorCalculus}
I.~Colombaro, J.~Font-Segura, and A.~Martinez, ``An introduction to space--time
  exterior calculus,'' \emph{Mathematics}, vol.~7, pp. 564--583, Jun. 2019.

\bibitem{colombaro2020generalizedMaxwellEquations}
------, ``Generalized {M}axwell equations for exterior-algebra multivectors in
  $(k,n)$ space-time dimensions,'' \emph{European Physical Journal Plus}, vol.
  135, no. 305, Mar. 2020.

\bibitem{lovelock1989tensorsDifferentialForms}
D.~Lovelock and H.~Rund, \emph{Tensors, Differential Forms, and Variational
  Principles}.\hskip 1em plus 0.5em minus 0.4em\relax Dover Publications, 1989.

\bibitem{flanders1989differentialForms}
H.~Flanders, \emph{Differential Forms with Applications to the Physical
  Sciences}.\hskip 1em plus 0.5em minus 0.4em\relax Dover Publications, 1989.

\bibitem{feynman1977lecturesPhysicsvol2}
R.~P. Feynman, R.~B. Leighton, and M.~Sands, \emph{The Feynman Lectures on
  Physics, Vol. II: Mainly Electromagnetism and Matter}.\hskip 1em plus 0.5em
  minus 0.4em\relax Addison-Wesley, 1977.

\bibitem{zee2003quantumFieldTheoryNutshell}
A.~Zee, \emph{{Quantum Field Theory in a Nutshell}}, ser. Nutshell
  handbook.\hskip 1em plus 0.5em minus 0.4em\relax Princeton, NJ: Princeton
  Univ. Press, 2003.

\bibitem{montesinos2006symmetricTensor}
M.~Montesinos and E.~Flores, ``Symmetric energy-momentum tensor in {M}axwell,
  {Y}ang-{M}ills, and {P}roca theories obtained using only {N}oether's
  theorem,'' https://arxiv.org/abs/hep-th/0602190.

\end{thebibliography}

\end{document}